\documentclass[titlepage,12pt,subeqn]{utarticle}
\usepackage{amsmath,amssymb,amscd}
\usepackage{latexsym}
\usepackage{hyperref}
\usepackage{ifthen}
\usepackage{bbm}

\usepackage{multirow}

\newcommand{\be}{\begin{equation}}
\newcommand{\ee}{\end{equation}}
\newcommand{\bea}{\begin{eqnarray}}
\newcommand{\eea}{\end{eqnarray}}
\newcommand{\al}{\alpha}
\renewcommand{\d}{\delta}
\newcommand{\e}{\epsilon}
\newcommand{\G}{\Gamma}
\newcommand{\g}{\gamma}
\renewcommand{\k}{\kappa}

\newcommand{\m}{\mu}

\newcommand{\Om}{\Omega}
\newcommand{\om}{\omega}
\newcommand{\s}{\sigma}
\renewcommand{\t}{\theta}
\newcommand{\C}{\mathbb{C}}

\newcommand{\hlf}{\frac{1}{2}}

\newcommand{\non}{\nonumber}

\newcommand{\p}{\partial}

\newcommand{\R}{\mathbb{R}}
\newcommand{\rr}{\rightarrow}
\newcommand{\w}{\wedge}
\newcommand{\Z}{\mathbb{Z}}
\newcommand{\GL}{\operatorname{GL}}
\renewcommand{\Im}{\operatorname{Im}}

\renewcommand{\Re}{\operatorname{Re}}

\newcommand{\SL}{\operatorname{SL}}
\newcommand{\SO}{\operatorname{SO}}

\newcommand{\SU}{\operatorname{SU}}
\newcommand{\U}{\operatorname{U}}
\newcommand{\lp}{\left(}
\newcommand{\rp}{\right)}
\newcommand{\ls}{\left[}
\newcommand{\rs}{\right]}
\numberwithin{equation}{section}

\begin{document}
\preprint{
UTTG-09-08 \\
MPP-2008-170\\
}

\title{On Slow-Roll Moduli Inflation in Massive\\[.5cm] IIA Supergravity with Metric Fluxes}

\author{ {\normalsize\sc Raphael Flauger}, {\normalsize\sc Sonia Paban}, {\normalsize\sc Daniel Robbins},
     \address{
      Theory Group, Department of Physics,\\
      University of Texas at Austin,\\
      Austin, TX 78712, USA\\
           }
      and {\normalsize\sc Timm Wrase}
      \address{
      Max-Planck-Institut f\"ur Physik,\\
      F\"ohringer Ring 6,\\
      80805 M\"unchen, Germany\\
      {~}\\
      {\rm Email}\\
       \emailt{flauger@physics.utexas.edu}\\
      \emailt{paban@zippy.ph.utexas.edu}\\
      \emailt{robbins@zippy.ph.utexas.edu}\\
      \emailt{wrase@mppmu.mpg.de}   }
      }



\Abstract{
We derive several no-go theorems in the context of massive type IIA string theory compactified to four dimensions in a way that, in the absence of fluxes, preserves $\mathcal{N}=1$ supersymmetry. Our derivation is based on the dilaton, K{\"a}hler and complex structure moduli dependence of the potential of the four-dimensional effective field theory, that is generated by the presence of D6-branes, O6-planes, RR fluxes, NSNS 3-form flux, and geometric fluxes. To demonstrate the usefulness of our theorems, we apply them to the most commonly studied class of toroidal orientifolds. We show that for all but two of the models in this class  the slow-roll parameter $\epsilon$ is bounded from below by numbers of order unity as long as the fluxes satisfy the Bianchi identities, ruling out slow-roll inflation and even the existence of de Sitter extrema in these models. For the two cases that avoid the no-go theorems, we provide some details of our numerical studies, demonstrating that small $\epsilon$ can indeed be achieved. We stress that there seems to be an $\eta$-problem, suggesting that none of the models in this class are viable from a cosmological point of view at least at large volume, small string coupling, and leading order in the $\alpha'$-expansion. 
}

\maketitle
\section{Introduction}
The increasing precision of cosmological data collected over the past two decades has  strengthened the case for inflation as the paradigm for early universe cosmology. Not only is the universe observed to be flat to within one per cent, the density perturbations are also found to be nearly scale-invariant, Gaussian, and adiabatic, just as predicted by inflation.
It may soon be possible to distinguish between various classes of inflationary models when upcoming experiments provide us with a more precise measurement of the power spectrum of scalar perturbations, measure small departures from Gaussianity or adiabaticity, or maybe even a B-mode signal, a pattern in the CMB polarization that would be the smoking gun of high-scale inflation.
According to the most naive estimates, the scale of inflation may only be few orders of magnitude below the Planck scale, implying that cosmology might even provide an experimental test for string theory \cite{Baumann:2008aq}, \cite{McAllister:2007bg}, quite likely the only one in the foreseeable future.

This opportunity has sparked work by many people toward an understanding of what string theory actually predicts in this regard, as well as work toward an understanding of the reliability of the approximations made to derive these predictions. Given that we do not understand why space-time has four large dimensions, the best we can do is to take this as an input and construct a reliable four dimensional effective theory, with stabilized moduli,  that produces inflation consistent with current observations in a way that allows us to control the corrections.
Most of the modern approaches are based on flux compactifications.  As explained in \cite{McAllister:2007bg}, one constructs a string inflation model by specifying the compactification manifold, its dimensionality and topology, the location of orientifold planes and D-branes, the type and amount of fluxes that are turned on and through which cycle. The size of the corrections will depend, among other things, on the value of  $g_s$, which controls the loop expansion, $\alpha' $, which sets the scale of the string modes, and the volume of the compactification manifold, which determines the mass scale for the Ka{\l}u\.{z}a-Klein modes.

Since the ground breaking work of KKLT~\cite{Kachru:2003aw}, which provided a mechanism to stabilize all moduli in the context of type IIB string theory, a lot of work has been done toward understanding inflation in string theory setups that are under control \cite{Quevedo:2002xw}, \cite{Linde:2005dd}, \cite{HenryTye:2006uv}, \cite{Cline:2006hu}, \cite{Burgess:2007pz}, \cite{Kallosh:2007ig}, \cite{McAllister:2007bg}. This effort has uncovered a variety of potential mechanisms for inflation which are usually broadly classified according to the origin of the inflaton field into moduli inflation or D-brane inflation.  A lot of progress has been made in the context of type IIB string theory; type IIA on the other hand still remains much less explored.

In type IIA,  work by  DeWolfe et al. \cite{DeWolfe:2005uu} gave an explicit construction for moduli stabilization relying on perturbative effects alone,  thus making it more reliable as far as calculability is concerned and allowing for more explicit constructions than in type IIB.  All is not good, however, since the presence of orientifolds in these compactifications may invalidate the usual effective field theory treatment as pointed out by Banks and van den Broek \cite{Banks:2006hg}. Although, we lack a good argument to appease those concerned by  the latter point, in this paper we will continue to ignore the backreaction of the orientifold plane in our search for inflation in massive type IIA, following Hertzberg et al.~\cite{Hertzberg:2007ke}, \cite{Hertzberg:2007wc}. The concerned reader may view our work as an attempt to derive some expertise and insight into the analysis of inflation in models with a large number of fields and fluxes that may also be useful in other contexts.

Generically, the various moduli fields are coupled to each other, and one has to search for slow-rolling regions in the typically rather high-dimensional moduli space.  Because of the large number of fields and the complexity that comes with it, the existence of a slow-roll path will typically either be excluded by analytical methods (no-go theorems) or will have to be confirmed by numerical analysis.

The original work of HTKS\"O  \cite{Hertzberg:2007ke} studied three simple orientifolds of $T^6$ \cite{DeWolfe:2005uu}, \cite{Villadoro:2005cu}, \cite{Ihl:2006pp} in the hope to find inflation, only to be disappointed, and HKTT  \cite{Hertzberg:2007wc} extended this work and proved a no-go theorem that applies to compactifications of IIA string theory on general Calabi-Yau manifolds with standard NSNS fluxes, RR fluxes, D6-branes, and O6-planes at large volume and small string coupling. This no-go theorem shows that the slow-roll parameter $\epsilon$ is bounded from below by some positive number independent of the choice of fluxes, implying that $\epsilon$ cannot be made small enough to allow for slow-roll inflation. This illustrates that even an infinite number of vacua does not guarantee that one of them will inflate.

As with any no-go theorem, the most important information contained in the theorem are the assumptions that went into its derivation. Beyond the usual assumptions of large volume, small string coupling, and leading order in the $\alpha'$-expansion, an important assumption in the case of HKTT was that the models did not include geometric or non-geometric fluxes.
In this work we  allow for geometric fluxes, but otherwise use the same assumptions as~\cite{Hertzberg:2007wc}. In addition, we assume that there is a hierarchy between the mass scales corresponding to the twisted and blow-up modes and the modes of the untwisted sector that we keep. For those D6-branes whose backreaction cannot be ignored, we limit ourselves to rigid embeddings, making a hierarchy between the open string modes and the modes we keep plausible so that they can be integrated out.\footnote{Phrased differently, the hierarchy guarantees that the branes will relax to their static configurations on time scales much shorter than the time scales we are interested in so that their dynamics can be ignored.} For branes that can consistently be treated as probes, this limitation does not apply. Under these assumptions, we derive several new no-go theorems,\footnote{As a caveat to the reader, we would like to note that our use of the phrase ``no-go theorem'' is probably an abuse of terminology for most of our cases because they apply to fairly narrow circumstances and require a relatively large number of assumptions (though we believe that our assumptions are reasonable). We chose this phrase for its familiarity to those who work in the subject and to evoke a certain set of connotations. However, any reader who is uncomfortable with this weakening of established terminology should feel free to substitute the alternative phrase ``no-go lemma'' at each occurence.} and, to demonstrate their usefulness, apply them to toroidal orientifolds with abelian orbifold groups generated by rotations and reflections, that, in the absence of fluxes and after orientifolding, preserve $\mathcal{N}=1$ supersymmetry. We show that in all these models except for the two $\mathbb{Z}_2\times\mathbb{Z}_2$ cases, even in the presence of geometric fluxes, the slow-roll parameter $\epsilon$ is bounded from below by numbers of order unity as long as the fluxes satisfy the Bianchi identities. For these two special cases, we numerically succeed in identifying regions in moduli space with arbitrarily small $\epsilon$, but all these regions seem to have too short a period of inflation. In other words, there is an $\eta$-problem similar to the one that has already appeared in many of  the type IIB compactifications, with the additional feature that at least one of these directions always turns out to be not only steep but also tachyonic.

While the original work of HKTT  \cite{Hertzberg:2007wc} was based exclusively on the dependence of the potential on the dilaton and volume moduli field our work is based on the dependence of the potential on the dilaton, volume as well as the K\"{a}hler and complex structure moduli.
No-go theorems of this kind are useful because they sharpen our understanding of the ingredients necessary for successful slow-roll inflation and allow us to exclude entire regions in the large landscape of solutions of string theory. The analysis of HKTT shows that localized sources such as NS5-branes or geometric or non-geometric fluxes are a necessary condition for slow-roll in type IIA string theory, we show that geometric fluxes alone are not generally sufficient.

There have by now been other studies of type IIA compactifications that have successfully identified regions of slow-roll inflation or found de Sitter vacua.

In~\cite{Silverstein:2007ac}, Silverstein constructed de Sitter solutions based on compactifications of type IIA on a product of two Nil-manifolds with orientifold planes, fivebranes, fractional Chern-Simons forms, and fluxes. As was shown by Silverstein and Westphal~\cite{Silverstein:2008sg}, in these setups it is not only possible to realize successful slow-roll inflation, but even to realize large-field inflation in a controlled way thus generating an observable tensor signal~\cite{Silverstein:2008sg}, making these the most exciting solutions in type IIA from a cosmological point of view so far.

While these vacua are presumably much closer to what one might expect a generic vacuum to look like, and one might argue that these are a natural place to start looking for inflation, they clearly are also somewhat less explicit and hence less controlled. So it seems worthwhile asking what the minimal set of ingredients for successful inflation or de Sitter solutions in type IIA is. Our studies are very much in this spirit, and it is also in this spirit that Haque et al.~\cite{Haque:2008jz} studied compactifications on a product of two maximally symmetric hyperbolic spaces. Another possible approach to evading the no-go theorem of~\cite{Hertzberg:2007wc}, complementary to the work here, is to appeal to corrections to the tree-level effective action. This is the direction pursued by the authors of~\cite{Palti:2008mg}.

Very recently, compactifications of type IIA on orientifolds of SU(3)-structure manifolds with non-vanishing geometric fluxes were studied in~\cite{Caviezel:2008tf}. Some of the models were ruled out in~\cite{Caviezel:2008tf} based on our results, for others numerical studies indicate that there are regions in moduli space with small $\epsilon$. However, the same $\eta$-problem seems to make an appearance, ruling these out from a cosmological perspective as well. This seems natural because the models studied there are in fact close cousins of the models that lead to small $\epsilon$ and an $\eta$-problem for us.

The organization of the paper is as follows. In section 2.1 and 2.2, we  review in some detail the low energy effective theory in the presence of both geometric and non-geometric fluxes. In section 2.3, we write explicit expressions for the slow-roll parameters $ \epsilon$ and $\eta$ as functions of the moduli and the fluxes. In section 3, we derive several no-go theorems based on the dependence of the effective potential on the dilaton, as well as K{\"a}hler and complex structure moduli. In section 4, we present a classification of the possible orientifolds of $T^6$ as well as the possible constraints on the fluxes, Bianchi identities and tadpoles conditions. In section 5, we apply the no-go theorems to the toroidal orientifolds discussed in section 4 and show that most of them cannot have small $\epsilon$, ruling out both slow-roll inflation and the existence of de Sitter extrema. In section 6, we present some details of our numerical studies for the models for which we found small $\epsilon$. We note that there seems to be an $\eta$ problem, implying that none of the models in this class can accommodate extended periods of inflation as seems to be required by observations.  We conclude in section 7. Appendix A contains a summary of our conventions.

\section{Low Energy Theory}

\subsection{Metric and non-geometric fluxes}
\label{subsec:generalized}

As mentioned above, recent work \cite{Hertzberg:2007wc} has shown that type IIA string theory with only ordinary fluxes ($H$-flux and RR fluxes) is not sufficient to allow for slow-roll inflation.  For this reason we would like to include some extra ingredients in an effort to overcome this obstacle.  There are many objects that one could add, and arguments that generically their presence will allow for de Sitter extrema \cite{Silverstein:2007ac}, but we would like to look instead for a minimal set of additional ingredients, and we will focus on one particular class of ingredients which are known as generalized NSNS fluxes.  Our motivation is that T-duality guarantees that these fluxes can appear (they should be on the same footing as ordinary $H$-flux), and T-duality also shows us how they must appear for consistency.

It is well-known that by T-dualizing a circle that is threaded by $H$-flux (that is if the circle isometry contracted with $H$ is non-zero), one obtains a new solution in which some components of $H$-flux have been exchanged for some non-constant metric components, whose effect can be thought of as a twist of the circle over the rest of the geometry.  These twists can be encoded in components $f^i_{jk}$, analogous to the individual components $H_{ijk}$ of the $H$-flux.  By performing an explicit Ka{\l}u\.{z}a-Klein reduction from ten dimensions, one can learn how such objects appear in the low-energy theory in four dimensions~\cite{Scherk:1979zr}, \cite{Kaloper:1998kr}, \cite{Kaloper:1999yr}.  It turns out that these objects, which are usually called metric fluxes (or sometimes geometric fluxes) because of the analogy with $H$, appear in the low-energy theory in much the same way that $H$ does.  If we started with an underlying space preserving $\mathcal{N}=1$, such as a Calabi-Yau orientifold of IIA, then the objects appear as parameters in the four-dimensional superpotential and the tadpole constraints.  It turns out that metric fluxes can also have the effect of giving a charge to some of the fields (fields that were RR axions before fluxes were turned on) under the four-dimensional vector multiplets, with the result that they can also appear in D-terms in four dimensions~\cite{Ihl:2007ah}, \cite{Robbins:2007yv}.

Sometimes there are further T-dualities one can perform, converting the metric flux components $f^i_{jk}$ into new objects $Q^{ij}_k$ known as non-geometric fluxes.  In the presence of these, the six-dimensional compact space is not a geometric manifold anymore, but rather it has the structure of a torus fiber glued over a base with transition functions that sit inside the full T-duality group~\cite{Hellerman:2002ax}, \cite{Dabholkar:2002sy}.  In fact, from a four-dimensional perspective, it is quite reasonable to expect a full set of these objects, $H_{ijk}$, $f^i_{jk}$, $Q^{ij}_k$, as well as objects labeled $R^{ijk}$, but not all of these have been explicitly constructed from a ten-dimensional string theory.  However, that doesn't necessarily stop us from discussing how they would appear in the low-energy theory, since that is determined by symmetry considerations~\cite{Shelton:2005cf}, \cite{Shelton:2006fd}, \cite{Aldazabal:2007sn}, \cite{Ihl:2007ah}, \cite{Robbins:2007yv}.  For a review of these constructions, the reader is encouraged to refer to the review~\cite{Wecht:2007wu}, and references therein.

Here, we will be satisfied to include a few more comments about metric fluxes.  Consider the case of $T^6$ in particular (this will be the starting point for all of our explicit constructions later).  In the presence of $f^i_{jk}$, the underlying geometry changes from a torus to a {\it twisted torus}, and the globally defined one-forms are no longer the closed forms $dx^i$, but rather forms $\eta^i$ that satisfy
\be
d\eta^i=-\hlf f^i_{jk}\eta^j\w\eta^k.
\ee
Simply demanding that $d^2=0$ gives us some constraints,
\be
f^i_{j[k}f^j_{\ell m]}=0.
\ee
Similarly, if we now expand $H$ in this basis, $H=\frac{1}{6}H_{ijk}\eta^i\w\eta^j\w\eta^k$, then the usual condition that $H$ be closed gives
\be
f^i_{[jk}H_{\ell m]i}=0.
\ee
These two sets of equations will be referred to as Bianchi identities, and the flux components we turn on must satisfy them for reasons of consistency.  Finally, it turns out to be much more convenient to express our fluxes instead using a basis of forms that, in the absence of metric fluxes, would be the
harmonic forms of the underlying space (see Appendix \ref{sec:Conventions} for our conventions),
\be
H=p_Kb^K,\qquad d\om_a=-r_{aK}b^K,\qquad d\mu_\al=-\hat{r}_\al^Ka_K.
\ee
Here $p_K$, $r_{aK}$, and $\hat{r}_\al^K$ are linear combinations of $H_{ijk}$ and $f^i_{jk}$.  More details can be found in section \ref{subsec:fluxes}.

Throughout this paper we will only turn on $H$-flux and metric fluxes, and not any of the non-geometric fluxes, since we generically expect that in the presence of non-geometric fluxes, some volume moduli will be stuck near the string scale (since the transition functions involve T-dualities that include volume inversions).  However, it will be useful to keep the non-geometric fluxes in our minds when thinking about using T-dualities to convert one configuration of fluxes into a more useful configuration, as discussed in section \ref{subsec:ressym}.

\subsection{Effective potential}
\label{sec:EffectivePotential}

Our starting point is a Calabi-Yau orientifold of type IIA string theory.  We will add RR fluxes, as well as $H$-flux and metric flux from the NSNS sector.  Our conventions are listed in Appendix \ref{sec:Conventions}.

These ingredients then lead to an effective $\mathcal{N}=1$ supergravity theory in four dimensions.  To describe this effective theory, and particularly the effective potential for the complex scalar fields $t^a$ and $N^K$, we must provide the K\"ahler potential $K$, the holomorphic superpotential $W$, the holomorphic gauge kinetic couplings $f_{\al\beta}$, and the gauge transformations of the scalar fields under the different $\U(1)$ gauge groups arising from the four-dimensional vectors (i.e. we must give the electric and magnetic charges of the scalar fields).  Then the effective action for the scalars is
\be
S=-\int\left\{K_{a\bar b}dt^a\w\ast d\overline{t^b}+K_{I\bar J}dN^I\w\ast d\overline{N^J}+V\ast 1\right\},
\ee
where the scalar potential is
\be
V=e^K\lp K^{a\bar b}D_aW\overline{D_bW}+K^{I\bar J}D_IW\overline{D_JW}-3|W|^2\rp+\hlf\lp\operatorname{Re}f\rp^{-1\,\al\beta}D_\al D_\beta.
\ee
Here, $\ast$ is the four-dimensional Hodge star, $K_{a\bar b}=\frac{\p}{\p t^a}\frac{\p}{\p\overline{t^b}}K$, $K^{a\bar b}$ is its (transpose) inverse, $D_aW=\frac{\p}{\p t^a}W+(\frac{\p}{\p t^a}K)W$, and similarly for the $N^K$, and the D-terms are
\be
D_\al=\frac{i}{W}\lp\d_\al t^aD_aW+\d_\al N^KD_KW\rp=i\lp\d_\al t^a\p_aK+\d_\al N^K\p_KK\rp+i\frac{\d_\al W}{W},
\ee
where $\lambda^\al\d_\al\phi$ is the variation of the field $\phi$ under an infinitesimal gauge transformation $A^\al\rr A^\al+d\lambda^\al$.  One can also discuss D-terms arising from the magnetic gauge groups, but the details are similar.

\noindent For the IIA orientifolds at hand, we can provide this information~\cite{Grimm:2004ua}.  The K\"ahler potential is given by\footnote{Here and in some formulae below we will sometimes drop certain indices in cases where the contractions are obvious.  For instance, $(\k v^3)$ will be short hand for $\k_{abc}v^av^bv^c$.  We hope that this does not cause the reader too much difficulty.}
\be
K=4D-\ln\lp\frac{4}{3}\lp\k v^3\rp\rp.
\ee

\noindent In the sector of K\"ahler moduli, this leads to
\be
\p_aK=\frac{3i}{2}\frac{\lp\k v^2\rp_a}{\lp\k v^3\rp},
\ee
and
\be
K_{a\bar b}=\frac{9}{4}\lp\k v^3\rp^{-2}\lp\k v^2\rp_a\lp\k v^2\rp_b-\frac{3}{2}\lp\k v^3\rp^{-1}\lp\k v\rp_{ab},
\ee
with inverse
\be
K^{a\bar b}=-\frac{2}{3}\lp\k v^3\rp\lp\k v\rp^{-1\,ab}+2v^av^b.
\ee
Note that there is a no-scale condition
\be
K^{a\bar b}\p_aK\overline{\p_bK}=3.
\ee

\noindent In the complex structure sector, our moduli are defined by $N^K = \hlf\xi^K+ie^{-D}\mathcal{Z}^K$, where $D$ is the four-dimensional dilaton and the $\mathcal{Z}^K$ come from the expansion of the holomorphic three-form $\Om$.  These $\mathcal{Z}^K$ are not all independent (there are $h^{2,1}+1$ of them which are functions of the $h^{2,1}$ complex structure moduli), and in fact they satisfy a relation which can always be written as
\be \label{eq:prepot}
p_n(\mathcal{Z})=1,
\ee
where $p_n(\mathcal{Z})$ is a homogeneous polynomial of degree $n=h^{2,1}+1$ in the $\mathcal{Z}^K$.  In terms of this polynomial (which of course plays the role of a prepotential) we can then write
\be
\mathcal{F}_K=-\frac{i}{2n}\frac{\p}{\p\mathcal{Z}^K}p_n(\mathcal{Z}),
\ee
and
\be
D=-\frac{1}{n}\ln\ls p_n(I)\rs,
\ee
where $I^K=\Im N^K=e^{-D}\mathcal{Z}^K$.  It then follows that
\be
\frac{\p}{\p N^J}K=\frac{2i}{n}\frac{\p_J p_n(I)}{p_n(I)}=\frac{2i}{n}e^D\p_J p_n(\mathcal{Z})=-4e^D\mathcal{F}_J,
\ee
\be
K_{J\bar{K}}=\frac{1}{n}e^{2D}\ls\p_Jp_n(\mathcal{Z})\p_Kp_n(\mathcal{Z})-\p_J\p_Kp_n(\mathcal{Z})\rs.
\ee
It can be useful to pull out the dilaton dependence here and define
\be
\widehat{K}_{JK}=e^{-2D}K_{J\bar{K}}=\frac{1}{n}\ls\p_Jp_n(\mathcal{Z})\p_Kp_n(\mathcal{Z})-\p_J\p_Kp_n(\mathcal{Z})\rs.
\ee
The inverse of $\widehat{K}_{JK}$ will simply be denoted by $\widehat{K}^{KL}$, so that $K^{KL}=e^{-2D}\widehat{K}^{KL}$.

\noindent With these results it is easy to verify the identities
\be
\widehat{K}_{JK}\mathcal{Z}^K=2i\mathcal{F}_J,\qquad\widehat{K}^{JK}\mathcal{F}_K=-\frac{i}{2}\mathcal{Z}^J,
\ee
\be
\mathcal{Z}^K\mathcal{F}_K=-\frac{i}{2},\qquad\widehat{K}^{JK}\mathcal{F}_J\mathcal{F}_K=-\frac{1}{4},\qquad\widehat{K}_{JK}\mathcal{Z}^J\mathcal{Z}^K=1,
\ee
and the no-scale-type condition
\be
K^{J\bar{K}}\p_JK\overline{\p_KK}=4.
\ee

\noindent The gauge kinetic couplings are
\be
f_{\al\beta}=i\lp\widehat{\k}t\rp_{\al\beta},
\ee
with
\be
\lp\operatorname{Re}f\rp^{-1}=-\lp\widehat{\k}v\rp^{-1}.
\ee
The corresponding D-terms are
\be
D_\al=2ie^D\lp\widehat{r}\mathcal{F}\rp_\al,
\ee
so that the D-term contribution to the potential is
\be
V_D = \hlf\lp\operatorname{Re}f\rp^{-1\,\al\beta}D_\al D_\beta= 2e^{2D}\lp\widehat{\k}v\rp^{-1}(\widehat{r}\mathcal{F})^2.
\ee

\noindent Unlike the K\"ahler potential, the superpotential depends on the fluxes,
\be
W=e_0+ te+\hlf\k t^2m+\frac{1}{6}\widetilde{m}\lp\k t^3\rp+2 Np+2 Nrt.
\ee
The covariant derivatives of $W$ are
\bea
D_aW &=&  e_a+\lp\k mt\rp_a+\hlf\widetilde{m}\lp\k t^2\rp_a+2\lp Nr\rp_a+\frac{3i}{2}\frac{\lp\k v^2\rp_a}{\lp\k v^3\rp}W,\non\\
D_KW &=& 2p_K+2\lp rt\rp_K-4e^D\mathcal{F}_KW.
\eea

\noindent We will also mention that there are tadpole conditions which the generalized fluxes should satisfy,
\be
-\sqrt{2}\lp p_K\widetilde{m}-r_{aK}m^a\rp=2N_K^{(\mathrm{O6})}-N_K^{(\mathrm{D6})}.
\ee
where the right hand side represents the contribution of localized sources, both O6-planes and D6-branes.

\subsection{Slow-roll parameters}

As discussed in~\cite{Hertzberg:2007ke}, we can express the slow-roll parameters $\e$ and $\eta$ in terms of the scalar potential and K\"ahler metric.

\noindent The expression for $\e$ is
\bea
\e &=& V^{-2}\left\{ K^{a\bar b}\frac{\p}{\p t^a}V\frac{\p}{\p\bar{t}^{\bar b}}V+K^{I\bar J}\frac{\p}{\p N^I}V\frac{\p}{\p\bar{N}^{\bar J}}V\right\}\non\\
&=& \frac{1}{4}V^{-2}\left\{ K^{a\bar b}\lp\frac{\p}{\p v^a}V\frac{\p}{\p v^b}V+\frac{\p}{\p u^a}V\frac{\p}{\p u^b}V\rp\right.\\
&& \qquad\left.+K^{I\bar J}\lp\frac{\p}{\p\Re N^I}V\frac{\p}{\p\Re N^J}V+\frac{\p}{\p\Im N^I}V\frac{\p}{\p\Im N^J}V\rp\right\}.\non
\eea

\noindent If we further define $v^a=\rho\g^a$, where
\be
\k_{abc}\g^a\g^b\g^c=6,
\ee
so that the over-all volume is $\mathcal{V}_6=\rho^3$, and use the expressions for the K\"ahler metric above, then we can further simplify this to
\bea\label{eq:epsilon}
\e &=& V^{-2}\left\{\frac{1}{3}\rho^2\lp\frac{\p V}{\p\rho}\rp^2+\frac{1}{4}\lp\frac{\p V}{\p D}\rp^2\right.\non\\
&& \qquad\left.+\ls-\lp\k\g\rp^{-1\,ab}+\frac{1}{6}\g^a\g^b\rs\frac{\p V}{\p\g^a}\frac{\p V}{\p\g^b}+\frac{1}{4}\ls\widehat{K}^{JK}-\mathcal{Z}^J\mathcal{Z}^K\rs\frac{\p V}{\p\mathcal{Z}^J}\frac{\p V}{\p\mathcal{Z}^K}\right.\non\\
&& \qquad\left.+\rho^2\ls-\lp\k\g\rp^{-1\,ab}+\frac{1}{2}\g^a\g^b\rs\frac{\p V}{\p u^a}\frac{\p V}{\p u^b}+e^{-2D}\widehat{K}^{JK}\frac{\p V}{\p\xi^J}\frac{\p V}{\p\xi^K}\right\}.
\eea
This expression splits $\e$ into six non-negative pieces.  The first line involves the overall volume modulus $\rho$ and the four-dimensional dilaton $D$.  In this class of models, the potential $V$ can always be written as a polynomial in $\rho$ and $e^D$, so this line is often very easy to compute.  The no-go theorems of~\cite{Hertzberg:2007wc} have been derived by focussing only on this line.

The second line involves the angular K\"ahler moduli, $\g^a$ and the complex structure moduli $\mathcal{Z}^J$.  Both these sets of variables are constrained ($\k\g^3=6$, $\widehat{K}\mathcal{Z}^2=1$), so there is no unique way of writing the potential in terms of them.  However, the metrics which appear in the expression above are such that $\e$ doesn't depend on these choices.  For example, if a $\k\g^3$ appears anywhere in $V$, then when the derivative with respect to $\g^a$ hits it we get a contribution proportional to $\k_{abc}\g^b\g^c$, but this is annihilated by the term in square brackets above.
Finally the third line contains the axions $u^a$ and $\xi^J$.

The expression for $\eta$ is slightly more complicated.  First we must define a canonical metric $g_{ij}$ on the moduli space of real fields, given by
\be
\hlf g_{ij}d\phi^id\phi^j=K_{A\bar{B}}d\Phi^Ad\overline{\Phi^B}.
\ee
Here we are using the indices $i$ and $j$ to run over all real valued fields, while $A$ and $B$ run over all complex-valued fields, from both the complex structure and K\"ahler sectors of the theory.
From this metric we can then compute Christoffel symbols $\G^i_{jk}$, and then we have
\be
\eta=\mathrm{minimum\ eigenvalue\ of}\ \left\{\frac{g^{ik}\lp\p_k\p_jV-\G^\ell_{kj}\p_\ell V\rp}{V}\right\}.
\ee

\subsection{Scalar potential with metric fluxes}

\noindent The full expression for the scalar potential in this case is given by
\bea
V &=& \frac{3}{\lp\k v^3\rp}e^{2D}\left\{\widehat{K}^{IJ}\lp p_I+r_{aI}u^a\rp\lp p_J+r_{bJ}u^b\rp+\widehat{K}^{IJ}r_{aI}r_{bJ}v^av^b-2\lp\mathcal{Z}^Kr_{aK}v^a\rp^2\right.\non\\
&& \qquad\left.-\frac{2}{3}\lp\k v^3\rp\lp\k v\rp^{-1\,ab}\mathcal{Z}^I\mathcal{Z}^Jr_{aI}r_{bJ}\right\}+2e^{2D}\lp \widehat{\k}v\rp^{-1\,\al\beta}\lp\widehat{r}_\al^I\mathcal{F}_I\rp\lp\widehat{r}_\beta^J\mathcal{F}_J\rp\non\\
&& \quad+2e^{3D}\mathcal{Z}^K\lp\widetilde{m}p_K-r_{aK}m^a\rp\\
&& \quad+\frac{3}{4\lp\k v^3\rp}e^{4D}\left\{\ls-\frac{2}{3}\lp\k v^3\rp\lp\k v\rp^{-1\,ab}+2v^av^b\rs\right.\non\\
&& \qquad\left.\times\lp\xi^Ir_{aI}+e_a+\lp\k mu\rp_a+\hlf\widetilde{m}\lp\k u^2\rp_a\rp\lp\xi^Jr_{bJ}+e_b+\lp\k mu\rp_b+\hlf\widetilde{m}\lp\k u^2\rp_b\rp\right.\non\\
&& \qquad\left.+4\ls\xi^K\lp p_K+r_{aK}u^a\rp+\widetilde{e}+\lp eu\rp+\hlf\lp\k mu^2\rp+\frac{1}{6}\widetilde{m}\lp\k u^3\rp\rs^2\right.\non\\
&& \qquad\left.+\ls-\frac{2}{3}\lp\k v^3\rp\lp\k v\rp_{ab}+\lp\k v^2\rp_a\lp\k v^2\rp_b\rs\lp m^a+\widetilde{m}u^a\rp\lp m^b+\widetilde{m}u^b\rp+\frac{1}{9}\widetilde{m}^2\lp\k v^3\rp^2\right\}.\non
\eea

\section{No-Go Theorems}

In this section we will prove a series of no-go theorems.  Each one will show that, given some restrictions on the model, the slow-roll parameter is bounded below by some positive number of order unity, thus ruling out both slow-roll inflation and the existence of de Sitter extrema.  There will be two types of restrictions that we will consider.  We might impose conditions on the intersection numbers of the model, as encoded by the polynomials $\k_{abc}$, $\widehat{\k}_{a\,\al\beta}$ and the polynomial $p_n$, and we will further restrict which fluxes can be turned on.

\subsection{General manifolds}

Consider first the case where no restrictions are assumed on the intersection numbers of the model, that is $\k$, $\widehat{\k}$ and $p_n$ are unconstrained.

Here we will demonstrate two no-go theorems.  The first was shown by~\cite{Hertzberg:2007wc} and pertains to the case of no metric fluxes, that is $r_{aK}=\widehat{r}_\al^K=0$.  In this case the scalar potential simplifies to
\bea
V &=& \hlf\rho^{-3}e^{2D}\widehat{K}^{IJ}p_Ip_J+2\widetilde{m}e^{3D}\mathcal{Z}^Kp_K\non\\
&& \quad+\frac{1}{8}\rho^{-3}e^{4D}\left\{4\rho^2\ls-\lp\k\g\rp^{-1\,ab}+\hlf\g^a\g^b\rs\right.\non\\
&& \qquad\left.\times\lp e_a+\lp\k mu\rp_a+\hlf\widetilde{m}\lp\k u^2\rp_a\rp\lp e_b+\lp\k mu\rp_b+\hlf\widetilde{m}\lp\k u^2\rp_b\rp\right.\\
&& \qquad\left.+\ls\xi^Kp_K+\widetilde{e}+\lp eu\rp+\hlf\lp\k mu^2\rp+\frac{1}{6}\widetilde{m}\lp\k u^3\rp\rs^2\right.\non\\
&& \qquad\left.+\rho^4\ls-4\lp\k\g\rp_{ab}+\lp\k\g^2\rp_a\lp\k\g^2\rp_b\rs\lp m^a+\widetilde{m}u^a\rp\lp m^b+\widetilde{m}u^b\rp+4\widetilde{m}^2\rho^6\right\}.\non
\eea
Note that the metrics $[-(\k\g)^{-1\,ab}+\hlf\g^a\g^b]$ and $[-4(\k\g)_{ab}+(\k\g^2)_a(\k\g^2)_b]$ are both positive definite since they are equal to $\frac{1}{4}\rho^{-2}K^{a\bar{b}}$ and $16\rho^2K_{a\bar{b}}$ respectively, so that the only term in the potential which can be negative is the second term on the first line above.

\noindent From this we can easily check that
\be
3\p_DV-\rho\p_\rho V\ge 9V.
\ee
Finally, if we also have $V>0$, then we can write
\be
\e\ge V^{-2}\ls\frac{1}{3}\rho^2\lp\p_\rho V\rp^2+\frac{1}{4}\lp\p_DV\rp^2\rs=V^{-2}\ls\frac{1}{39}\lp 3\p_DV-\rho\p_\rho V\rp^2+\frac{1}{52}\lp\p_D+4\rho\p_\rho V\rp^2\rs\ge\frac{27}{13}.
\ee

\noindent Let us consider another example of possible interest, where we allow metric fluxes, but don't allow a Romans mass parameter, that is we take $\widetilde{m}=0$.  This would be the models one would look at if one wished to have a straightforward lift to M-theory, for example.  In this case one can easily check that there is another no-go theorem.  Indeed, we have
\be
\p_DV-\rho\p_\rho V\ge 3V,
\ee
and so
\be
\e\ge V^{-2}\ls\frac{1}{7}\lp\p_DV-\rho\p_\rho V\rp^2+\frac{1}{84}\lp3\p_DV+4\rho\p_\rho V\rp^2\rs\ge\frac{9}{7}.
\ee

\noindent Thus in both these cases there is no possibility of slow-roll inflation, and no possibility of finding a de Sitter extremum of the potential anywhere in field space (such a point would have $\e=0$, of course).  For this reason we will assume that $\widetilde{m}\ne 0$ in subsequent sections, and we will focus on cases in which some metric fluxes are non-zero.
The observation that a non-zero $\widetilde{m}$ is necessary for the existence of de Sitter vacua was made independently in~\cite{Haque:2008jz}.

\subsection{Factorization in the K\"ahler sector}

Now consider a more restricted class of models, in which there is one distinguished K\"ahler modulus $v^0$, such that the only non-zero intersections are
\be
\k_{0ij}=X_{ij},\qquad\widehat{\k}_{0\,\al\beta}=\widehat{X}_{\al\beta},
\ee
and their permutations, where $i$ and $j$ run over the remaining K\"ahler moduli.  When considering general Calabi-Yau orientifolds, this is a very unnatural condition.  However, frequently one is interested in orientifolds of $T^6$ and there is a hierarchy between the moduli of the untwisted sector and those of the twisted sectors.  In such a case one typically truncates to the untwisted moduli, and in this sector the constraints above on the intersection numbers are not uncommon; they correspond to the presence of a $T^2$ factor in the $T^6$ which is preserved by the orientifold group.

In the general case, we found it profitable to split the K\"ahler moduli into an overall volume variable $\rho$ and a set of angular variables $\g^a$.  In the factorized case it is more useful to take $v^0$ and then split the remaining moduli by defining $v^i=\s\chi^i$, where the angular variables $\chi^i$ are constrained by
\be
X_{ij}\chi^i\chi^j=2.
\ee
Then, for instance, the volume of the space is $\mathcal{V}_6=v^0\s^2$.  With these conventions we find that the K\"ahler metric and its inverse have the form
\be
K_{a\bar{b}}=\lp\begin{matrix}\frac{1}{4\lp v^0\rp^2} & 0 \\ 0 & \frac{1}{4\s^2}\ls\lp X\chi\rp_i\lp X\chi\rp_j-X_{ij}\rs\end{matrix}\rp,\qquad K^{a\bar{b}}=\lp\begin{matrix}4\lp v^0\rp^2 & 0 \\ 0 & 4\s^2\ls-X^{-1\,ij}+\chi^i\chi^j\rs\end{matrix}\rp.
\ee

\noindent The combinations that appear in the expression for $\e$ are
\be
\frac{1}{4}K^{a\bar{b}}\frac{\p V}{\p v^a}\frac{\p V}{\p v^b}=\lp v^0\rp^2\lp\frac{\p V}{\p v^0}\rp^2+\hlf\s^2\lp\frac{\p V}{\p\s}\rp^2+\ls-X^{-1\,ij}+\hlf\chi^i\chi^j\rs\frac{\p V}{\p\chi^i}\frac{\p V}{\p\chi^j}.
\ee

\noindent We can now look for no-go theorems involving the three variables $D$, $v^0$, and $\s$.  For example, suppose $r_{0K}=0$, but we allow non-zero $r_{iK}$ and $\widehat{r}_\al^K$, then
\bea
V &=& \frac{1}{2v^0\s^2}e^{2D}\left\{\widehat{K}^{IJ}\lp p_I+r_{iI}u^i\rp\lp p_J+r_{jJ}u^j\rp+\s^2\widehat{K}^{IJ}r_{iI}r_{jJ}\chi^i\chi^j-4\s^2\mathcal{Z}^I\mathcal{Z}^JX^{-1\,ij}r_{iI}r_{jJ}\right\}\non\\
&& +2\lp v^0\rp^{-1}e^{2D}\widehat{X}^{-1\,\al\beta}\lp\widehat{r}_\al^I\mathcal{F}_I\rp\lp\widehat{r}_\beta^J\mathcal{F}_J\rp+2e^{3D}\mathcal{Z}^K\lp\widetilde{m}p_K-r_{iK}m^i\rp\non\\
&& +\frac{1}{2v^0\s^2}e^{4D}\left\{\lp v^0\rp^2\lp e_0+X_{ij}m^iu^j+\hlf\widetilde{m}X_{ij}u^iu^j\rp^2\right.\\
&& \left.+\s^2\ls-X^{-1\,ij}+\chi^i\chi^j\rs\lp\xi^Ir_{iI}+e_i+m^0X_{ik}u^k+u^0X_{ik}m^k+\widetilde{m}u^0X_{ik}u^k\rp\right.\non\\
&& \qquad\left.\times\lp\xi^Jr_{jJ}+e_j+m^0X_{jl}u^l+u^0X_{jl}m^l+\widetilde{m}u^0X_{jl}u^l\rp\right.\non\\
&& \left.+\ls\xi^K\lp p_K+r_{iK}u^i\rp+\widetilde{e}+e_0u^0+e_iu^i+\hlf\lp m^0+\widetilde{m}u^0\rp X_{ij}u^iu^j+u^0X_{ij}m^iu^j\rs^2\right.\non\\
&& \left.+\s^4\lp m^0+\widetilde{m}u^0\rp^2+\lp v^0\rp^2\s^2\ls-X_{ij}+X_{ik}X_{jl}\chi^k\chi^l\rs\lp m^i+\widetilde{m}u^i\rp\lp m^j+\widetilde{m}u^j\rp\right.\non\\
&& \left.+\widetilde{m}^2\lp v^0\rp^2\s^4\vphantom{\lp\hlf\rp^2}\right\}.\non
\eea
In this case it is easy to check that
\be
\p_DV-v^0\p_{v^0}V\ge 3V,
\ee
so that
\be
\e\ge V^{-2}\ls\frac{1}{5}\lp\p_DV-v^0\p_{v^0}V\rp^2+\frac{1}{20}\lp\p_DV+4v^0\p_{v^0}V\rp^2\rs\ge\frac{9}{5}.
\ee

\noindent Similarly, if $r_{iK}=\widehat{r}_\al^K=0$, but we have arbitrary $r_{0K}$, the potential has the form
\bea
V &=& \frac{1}{2v^0\s^2}e^{2D}\left\{\widehat{K}^{IJ}\lp p_I+r_{0I}u^0\rp\lp p_J+r_{0J}u^0\rp+\lp v^0\rp^2\widehat{K}^{IJ}r_{0I}r_{0J}\right\}\non\\
&& +2e^{3D}\mathcal{Z}^K\lp\widetilde{m}p_K-r_{0K}m^0\rp\non\\
&& +\frac{1}{2v^0\s^2}e^{4D}\left\{\lp v^0\rp^2\lp\xi^Kr_{0K}+e_0+X_{ij}m^iu^j+\hlf\widetilde{m}X_{ij}u^iu^j\rp^2\right.\non\\
&& \left.+\s^2\ls-X^{-1\,ij}+\chi^i\chi^j\rs\lp e_i+m^0X_{ik}u^k+u^0X_{ik}m^k+\widetilde{m}u^0X_{ik}u^k\rp\right.\non\\
&& \qquad\left.\times\lp e_j+m^0X_{jl}u^l+u^0X_{jl}m^l+\widetilde{m}u^0X_{jl}u^l\rp\right.\non\\
&& \left.+\ls\xi^K\lp p_K+r_{0K}u^0\rp+\widetilde{e}+e_0u^0+e_iu^i+\hlf\lp m^0+\widetilde{m}u^0\rp X_{ij}u^iu^j+u^0X_{ij}m^iu^j\rs^2\right.\non\\
&& \left.+\s^4\lp m^0+\widetilde{m}u^0\rp^2+\lp v^0\rp^2\s^2\ls-X_{ij}+X_{ik}X_{jl}\chi^k\chi^l\rs\lp m^i+\widetilde{m}u^i\rp\lp m^j+\widetilde{m}u^j\rp\right.\non\\
&& \left.+\widetilde{m}^2\lp v^0\rp^2\s^4\vphantom{\lp\hlf\rp^2}\right\},
\eea
and one can check that
\be
2\p_DV-\s\p_\s V\ge 6V,
\ee
giving
\be
\e\ge V^{-2}\ls\frac{1}{18}\lp 2\p_DV-\s\p_\s V\rp^2+\frac{1}{36}\lp\p_DV+4\s\p_\s V\rp^2\rs\ge 2.
\ee

\noindent Thus in order to get slow-roll inflation, we must have non-zero metric fluxes both with a $0$ index and without, where by fluxes without a $0$ index we mean either $r_{iK}$ or $\hat{r}_\al^K$.

\subsection{Factorization in the complex structure sector}

It is possible to find a similar sort of factorization in the complex structure sector.  Recall that the computations of the K\"ahler potential in this sector was determined by a polynomial $p_n$ which is homogeneous of degree $n=h^{2,1}+1$ in $n$ variables.  We defined the usual dilaton $D$ by writing $I^K=\Im(N^K)=e^{-D}\mathcal{Z}^K$, where the $\mathcal{Z}^K$ were constrained by $p_n(\mathcal{Z})=1$, or alternatively, $p_n(I)=e^{-nD}$.

Now suppose that we can divide the $I^K$ into two sets, $I_{(1)}^A$ and $I_{(2)}^P$ (we will use letters from different parts of the alphabet for the different sets), and that the polynomial $p_n$ factorizes as
\be
p_n(I)=p^{(1)}_{n_1}(I_{(1)})\cdot p^{(2)}_{n_2}(I_{(2)}),
\ee
where $n_1$ and $n_2$ are the degrees of the polynomials and satisfy $n_1+n_2=n$.\footnote{Note that the degrees $n_1$ and $n_2$ do not have to correspond to the cardinality of the sets $I_{(1)}^A$ and $I_{(2)}^P$.}  In this case we can define two dilatons, $D_1$ and $D_2$ by
\be
e^{-n_1D_1}=p^{(1)}_{n_1}(I_{(1)}),\qquad e^{-n_2D_2}=p^{(2)}_{n_2}(I_{(2)}),
\ee
and we can define two sets of $\mathcal{Z}$ by $\mathcal{Z}_{(1)}^A=e^{D_1}I_{(1)}^A$ and $\mathcal{Z}_{(2)}^P=e^{D_2}I_{(2)}^P$.  Each of these sets will be constrained, since $1=p^{(1)}_{n_1}(\mathcal{Z}_{(1)})=p^{(2)}_{n_2}(\mathcal{Z}_{(2)})$.  The K\"ahler potential in this sector is now given by
\be
K=-\frac{4}{n}\ln\ls p_n(I)\rs=4\lp\frac{n_1}{n}D_1+\frac{n_2}{n}D_2\rp.
\ee

The K\"ahler metric will then be block diagonal, with non-zero entries
\bea
K_{A\bar{B}} &=& e^{2D_1}\widehat{K}_{(1)\,AB}=e^{2D_1}\frac{1}{n}\ls\p_Ap^{(1)}_{n_1}(\mathcal{Z}_{(1)})\p_Bp^{(1)}_{n_1}(\mathcal{Z}_{(1)})-\p_A\p_Bp^{(1)}_{n_1}(\mathcal{Z}_{(1)})\rs,\\
K_{P\bar{Q}} &=& e^{2D_2}\widehat{K}_{(2)\,PQ}=e^{2D_2}\frac{1}{n}\ls\p_Pp^{(2)}_{n_2}(\mathcal{Z}_{(2)})\p_Qp^{(2)}_{n_2}(\mathcal{Z}_{(2)})-\p_P\p_Qp^{(2)}_{n_2}(\mathcal{Z}_{(2)})\rs.\non
\eea
Furthermore, in $\e$ we will find the combination
\bea
\frac{1}{4}K^{J\bar{K}}\frac{\p V}{\p I^J}\frac{\p V}{\p I^K} &=& \frac{n}{4n_1}\lp\frac{\p V}{\p D_1}\rp^2+\frac{n}{4n_2}\lp\frac{\p V}{\p D_2}\rp^2+\frac{1}{4}\ls\widehat{K}_{(1)}^{AB}-\frac{n}{n_1}\mathcal{Z}_{(1)}^A\mathcal{Z}_{(1)}^B\rs\frac{\p V}{\p\mathcal{Z}_{(1)}^A}\frac{\p V}{\p\mathcal{Z}_{(1)}^B}\non\\
&& \qquad+\frac{1}{4}\ls\widehat{K}_{(2)}^{PQ}-\frac{n}{n_2}\mathcal{Z}_{(2)}^P\mathcal{Z}_{(2)}^Q\rs\frac{\p V}{\p\mathcal{Z}_{(2)}^P}\frac{\p V}{\p\mathcal{Z}_{(2)}^Q}.
\eea

\noindent We can now try to concoct more no-go theorems working with the variables $D_1$, $D_2$, and $\rho$.  However, it turns out that we only gain an advantage over the general case if the non-zero flux contributions to the six-brane tadpoles come from only one of the subsets above, say only the subset labeled $(1)$.  In other words, we need to demand that $\widetilde{m}p_P=r_{aP}m^a$ (so that the tadpole contributions with a $P$-index vanish), while we allow $\widetilde{m}p_A-r_{aA}m^a\ne 0$.  In this case, and under certain extra conditions on the fluxes, we can find no-go theorems.  We present three examples, but the list is not exhaustive.

If $r_{aP}=\widehat{r}_\al^A=0$ and $n_1\ge n_2$, then we can show that $\p_{D_2}V\ge\frac{4n_2}{n}V$, which gives $\e\ge\frac{4n_2}{n}$.  Note in this case that combining $r_{aP}=0$ with our assumptions about the tadpoles forces $p_P=0$.

If $r_{aA}=\widehat{r}_\al^P=0$ then we have a family of inequalities of the form $3\p_{D_1}V+x\p_{D_2}V-\rho\p_\rho V\ge\frac{9n_1+(4x-3)n_2}{n}V$, where $x$ is any real number satisfying inequalities
\be
x\le 2,\qquad x\le \frac{5}{2}-\frac{n_1}{2n_2},\qquad x>\frac{3}{4}-\frac{9n_1}{4n_2}.
\ee
There are always solutions for $x$ and $\e$ turns out to always be maximized by taking $x$ at the top of the allowed interval, which leads to
\be
\begin{array}{ll}\e\ge\frac{49n_2}{n_1+28n_2}, & n_1\ge n_2, \\ \e\ge\frac{\lp 9n_1+5n_2\rp^2}{n\lp 39n_1+19n_2\rp}, & n_1\le n_2.\end{array}
\ee

Similarly, if $r_{aK}=0$ (i.e. both $r_{aA}$ and $r_{aP}$ vanish, and hence so also does $p_P$) then we can show that $3\p_{D_1}V+x\p_{D_2}V-\rho\p_\rho V\ge \frac{9n_1+(4x-3)n_2}{n}V$, where now $x$ is a real number satisfying
\be
x>\frac{3}{4}-\frac{9n_1}{4n_2},\qquad x\le\frac{5}{2}-\frac{n_1}{2n_2},\qquad\lp n_1-n_2\rp x\ge 4n_1-2n_2.
\ee
In this case there are solutions only when $5n_2>9n_1$, in which case the strongest bound is
\be
\e\ge\frac{\lp 9n_1-5n_2\rp^2}{39n_1^2-50n_1n_2+19n_2^2}.
\ee

\subsection{Factorization in both sectors}

Now, finally, let us briefly consider the case where there is factorization in both the K\"ahler and complex structure sectors of the theory, with the notation of the previous two sections.  There are many possible no-go theorems which can be derived in various situations.  One situation is relevant for our analysis below, so we present the derivation here.

This case occurs when we have $r_{0K}=0$ (that is both $r_{0A}=r_{0P}=0$) and $r_{iP}=p_P=\hat{r}^A=0$, but allow non-zero $r_{iA}$, $p_A$, and $\hat{r}^P$.  Here we find a family of inequalities,
$\p_{D_1}V+x\p_{D_2}V-v^0\p_{v^0}V\ge\frac{3n_1+(4x-1)n_2}{n}V$, where the real number $x$ must satisfy
\be
x>\frac{1}{4}-\frac{3n_1}{4n_2},\qquad\lp n_1-n_2\rp x\ge n_1-n_2.
\ee
These always admit solutions for $x$, and the corresponding bound on $\e$ is given by
\be
\begin{array}{ll}\e\ge\frac{9n_1+5n_2}{5n_1+n_2}, & n_1\ge n_2, \\ \e\ge\frac{9}{5}, & n_1<n_2.\end{array}
\ee

\section{Toroidal Orientifolds}

For a generic Calabi-Yau three-fold, it is not presently understood exactly how to consistently include metric fluxes or non-geometric fluxes.  When the manifold is at a point in its moduli space that admits a description as an orbifold of $T^6$, however, we can identify a subset of these generalized fluxes which can be turned on simply by twisting the torus construction.  In this case we can derive the full set of consistency conditions which must be satisfied, and it is for this reason that toroidal orbifolds and orientifolds are the most well-studied compactifications with generalized fluxes.

What will follow in the next section is a (partial) classification of orientifolds of $T^6$ which preserve $\mathcal{N}=1$ supersymmetry in four dimensions (see also a
related classification in \cite{Grana:2006kf}).  The goal is simply to generate a list of examples in which to look for slow-roll inflation or to test the utility of our no-go theorems.

\subsection{Classification of orientifolds}

There is a well-known classification of abelian orbifold groups which act on $T^6$ without shifts and which preserve $\mathcal{N}=2$ supersymmetry~\cite{Erler:1992ki}, \cite{Reffert:2006du}, \cite{Reffert:2007im}.  We will not be too concerned with the explicit action on the lattice, except in some specific cases in section \ref{sec:smalleexamples}.  As we will see, the action on the lattice only enters the story for us once we attempt to derive the correct quantization conditions on the generalized fluxes, but there is a great deal of information which can be obtained without these details.  With this in mind, then, we have nine different cyclic groups and eight more products of cyclic groups which can occur as lattice-preserving subgroups of $\SU(3)$, and hence give rise to $\mathcal{N}=2$ orbifolds of $T^6$.  Moreover, this list exhausts the possibilities for abelian orbifold groups, up to isomorphism.  These groups are listed in tables \ref{table:ZNActions} and \ref{table:ZNZMActions}.  In each table, the generator of the orbifold action is written as $\frac{1}{N}(n_1,n_2,n_3)$, which is shorthand for
\be
\lp z_1,z_2,z_3\rp\mapsto\lp e^{2\pi in_1/N}z_1,e^{2\pi in_2/N}z_2,e^{2\pi in_3/N}z_3\rp.
\ee

\begin{table}
\caption{Cyclic orbifold groups}
\centering
\begin{tabular}{|c|c|}
\hline Group $\Z_N$ & Generator $\frac{1}{N}\lp n_1,n_2,n_3\rp$ \\ \hline\hline$\Z_3$ & $\frac{1}{3}\lp 1,1,1\rp$ \\ \hline$\Z_4$ & $\frac{1}{4}\lp 1,1,2\rp$ \\ \hline$\Z_{6-I}$ & $\frac{1}{6}\lp 1,1,4\rp$ \\ \hline$\Z_{6-II}$ & $\frac{1}{6}\lp 1,2,3\rp$ \\ \hline$\Z_7$ & $\frac{1}{7}\lp 1,2,4\rp$ \\ \hline$\Z_{8-I}$ & $\frac{1}{8}\lp 1,2,5\rp$ \\ \hline$\Z_{8-II}$ & $\frac{1}{8}\lp 1,3,4\rp$ \\ \hline$\Z_{12-I}$ & $\frac{1}{12}\lp 1,4,7\rp$ \\ \hline$\Z_{12-II}$ & $\frac{1}{12}\lp 1,5,6\rp$ \\ \hline\end{tabular}
\label{table:ZNActions}
\end{table}

\begin{table}
\caption{$\Z_N\times\Z_M$ orbifold groups}
\centering
\begin{tabular}{|c|c|c|}
\hline Group $\Z_N\times\Z_M$ & First generator $\frac{1}{N}\lp n_1,n_2,n_3\rp$ & Second Generator $\frac{1}{M}\lp m_1,m_2,m_3\rp$ \\ \hline\hline$\Z_2\times\Z_2$ & $\frac{1}{2}\lp 1,0,1\rp$ & $\frac{1}{2}\lp 0,1,1\rp$ \\ \hline$\Z_2\times\Z_4$ & $\frac{1}{2}\lp 1,0,1\rp$ & $\frac{1}{4}\lp 0,1,3\rp$ \\ \hline$\Z_2\times\Z_6$ & $\frac{1}{2}\lp 1,0,1\rp$ & $\frac{1}{6}\lp 0,1,5\rp$ \\ \hline$\Z_2\times\Z_6'$ & $\frac{1}{2}\lp 1,0,1\rp$ & $\frac{1}{6}\lp 1,1,4\rp$ \\ \hline$\Z_3\times\Z_3$ & $\frac{1}{3}\lp 1,0,2\rp$ & $\frac{1}{3}\lp 0,1,2\rp$ \\ \hline$\Z_3\times\Z_6$ & $\frac{1}{3}\lp 1,0,2\rp$ & $\frac{1}{6}\lp 0,1,5\rp$ \\ \hline$\Z_4\times\Z_4$ & $\frac{1}{4}\lp 1,0,3\rp$ & $\frac{1}{4}\lp 0,1,3\rp$ \\ \hline$\Z_6\times\Z_6$ & $\frac{1}{6}\lp 1,0,5\rp$ & $\frac{1}{6}\lp 0,1,5\rp$ \\ \hline\end{tabular}
\label{table:ZNZMActions}
\end{table}

Let us now classify the supersymmetric orientifolds of these models.  An orientifold will be a $\Z_2$ extension $\widehat{G}$ of the orbifold group $G$,
\be
1\longrightarrow G\longrightarrow\widehat{G}\longrightarrow\Z_2\longrightarrow 1,
\ee
where each element of $\widehat{G}$ which is not in the image of $G$ (or equivalently not in the kernel of the map to $\Z_2$) must be accompanied by orientation reversal $\Omega$ and $(-1)^F$.  In order to preserve $\mathcal{N}=1$ supersymmetry in type IIA, we also require not only that the elements of $G$ act as linear holomorphic maps, $G\subset\SU(3)$, but we also demand that the orientation-reversing elements of $\widehat{G}$ act as linear {\it antiholomorphic} maps.  Thus for our classification we would like to find, for each of the orbifold group actions in the list above, all $\Z_2$ extensions, where the extension acts antiholomorphically.  To be more explicit, we want to find an antiholomorphic linear map $\s$ such that for every element $g$ in the orbifold group $G$, we have $g\s g\s\in G$.  As a set then, $\widehat{G}=G\cup \s G$.  In given holomorphic coordinates $z_1$, $z_2$, $z_3$, we can write each element as a three-by-three complex matrix with entries $g^i\vphantom{g}_j$, $\s^{\bar\imath}\vphantom{\s}_j$, and then the condition above is
\be
g^i\vphantom{g}_k{\bar\s}^k\vphantom{\bar\s}_{\bar\ell}{\bar g}^{\bar\ell}\vphantom{\bar g}_{\bar m}\s^{\bar m}\vphantom{\s}_j=\lp g'\rp^i\vphantom{\lp g\rp}_j,
\ee
for some $g'\in G$.

In all of the examples we will consider, the elements of $G$ are all diagonal with entries $g^i_j=\d^i_j\exp[i\t_i]$ (for instance see the generators in tables \ref{table:ZNActions} and~\ref{table:ZNZMActions}), so we have
\be
\label{eq:Extension}
\sum_{\bar k}e^{i(\t_i-\t_k)}{\bar\s}^i\vphantom{\bar\s}_{\bar k}\s^{\bar k}\vphantom{\s}_j=\d^i_je^{i\t'_i},
\ee
with no sum over $i$.

\noindent We will now consider different cases.

First consider the cases of cyclic groups whose generators have $\t_1$, $\t_2$, $\t_3$ all distinct (the last six cases in table \ref{table:ZNActions}).  In this case we can show that (\ref{eq:Extension}) requires ${\bar\s}^i\vphantom{\bar\s}_{\bar k}\s^{\bar k}\vphantom{\s}_j=0$ for all $\bar k$ (no sum on $\bar k$ here) and all $i\ne j$.  These equations can be shown to imply that only three entries of $\s$ are non-zero; either $\s$ is diagonal, or it is block diagonal with a one-by-one block and a two-by-two block with zeros on the diagonal.  In the diagonal case, we can apply phase changes to our holomorphic coordinates in order to transform $\s$ into the three-by-three identity matrix, so that it acts by simple conjugation, $z_i\mapsto\bar{z_i}$.  It turns out that this choice for $\s$ will be valid for each of our orbifold groups, and so we will denote it as the {\it standard orientifold} for each case.  These are summarized in table \ref{table:standardorientifolds}.  In the non-diagonal cases, we can again use a phase rotation to set the one-by-one block to one, and we can set one of the non-zero entries of the two-by-two block to one.  Then demanding that (\ref{eq:Extension}) be satisfied for each element of the orbifold group restricts the possibilities.  We find that there are no allowed non-standard orientifolds for $\Z_{6-II}$ and $\Z_7$, one choice for each of the $\Z_8$ groups, and two choices each for the $\Z_{12}$ groups, where to correctly count the number of independent choices, we should also recall that we can relabel our element $\s$ as $\s'=g\s$, for any $g\in G$, and then drop the prime.

\begin{table}
\caption{Standard $\mathcal{N}=1$ Orientifolds ($z_i\mapsto\bar{z_i}$)}
\centering
\begin{tabular}{|c|c|c|c|}
\hline Group & $h^{1,1}_{-\,\mathrm{untw}}$ & $h^{1,1}_{+\,\mathrm{untw}}$ & $h^{2,1}_\mathrm{untw}$ \\ \hline\hline$\Z_3$ & 6 & 3 & 0 \\ \hline$\Z_4$ & 4 & 1 & 1 \\ \hline$\Z_{6-I}$ & 4 & 1 & 0 \\ \hline$\Z_{6-II}$ & 3 & 0 & 1 \\ \hline$\Z_7$ & 3 & 0 & 0 \\ \hline$\Z_{8-I}$ & 3 & 0 & 0 \\ \hline$\Z_{8-II}$ & 3 & 0 & 1 \\ \hline$\Z_{12-I}$ & 3 & 0 & 0 \\ \hline$\Z_{12-II}$ & 3 & 0 & 1 \\ \hline$\Z_2\times\Z_2$ & 3 & 0 & 3 \\ \hline$\Z_2\times\Z_4$ & 3 & 0 & 1 \\ \hline$\Z_2\times\Z_6$ & 3 & 0 & 1 \\ \hline$\Z_2\times\Z_6'$ & 3 & 0 & 0 \\ \hline$\Z_3\times\Z_3$ & 3 & 0 & 0 \\ \hline$\Z_3\times\Z_6$ & 3 & 0 & 0 \\ \hline$\Z_4\times\Z_4$ & 3 & 0 & 0 \\ \hline$\Z_6\times\Z_6$ & 3 & 0 & 0 \\ \hline\end{tabular}
\label{table:standardorientifolds}
\end{table}

Next let us consider $\Z_4$ and $\Z_{6-I}$.  In this case we can easily show that $\s$ must be block diagonal with a two-by-two block for $z_1$ and $z_2$, and a one-by-one block for $z_3$.  A phase rotation can be used to set the latter entry to one (i.e. $\s\cdot z_3=\bar{z_3}$), but we have quite a bit more symmetries at our disposal in the two-by-two block, since any $\GL(2,\C)$ matrix commutes with the orbifold group.  It turns out that solving the constraints and then using the symmetries allows us to put $\s$ into one of two canonical forms.  Either we can set the two-by-two block to the identity, giving the standard orientifold, or we can set it to be the canonical antisymmetric matrix, so that $\s\cdot(z_1,z_2)=(\bar{z_2},-\bar{z_1})$.

In the case of $\Z_3$, we have $\t_1=\t_2=\t_3=2\pi i/3$.  Here (\ref{eq:Extension}) is the least restrictive, but we also have the most symmetry, since the full $\GL(3,\C)$ commutes with the orbifold group.  Here we can use this symmetry to always convert to the standard case.

We move on now to the product groups of table \ref{table:ZNZMActions}.  As in the case of the cyclic groups with distinct angles, we can show that $\s$ must be either diagonal, leading to the standard case, or block diagonal, with the two-by-two block having vanishing diagonal entries.  Finally, by carefully examining the remaining constraints and symmetries we are able to find the independent non-standard orientifolds in each case.  All of the non-standard orientifolds are summarized in table \ref{table:nonstandardorientifolds}.

\begin{table}
\caption{Non-Standard $\mathcal{N}=1$ Orientifolds}
\centering
\vskip .1cm
\begin{tabular}{|c|c|c|c|c|}
\hline Group & $\s\cdot\lp z_1,z_2,z_3\rp$ & $h^{1,1}_{-\,\mathrm{untw}}$ & $h^{1,1}_{+\,\mathrm{untw}}$ & $h^{2,1}_\mathrm{untw}$ \\ \hline\hline$\Z_4$ & $\lp\bar{z_2},-\bar{z_1},\bar{z_3}\rp$ & 2 & 3 & 1 \\ \hline$\Z_{6-I}$ & $\lp\bar{z_2},-\bar{z_1},\bar{z_3}\rp$ & 2 & 3 & 0 \\ \hline$\Z_{8-I}$ & $\lp\bar{z_3},\bar{z_2},\bar{z_1}\rp$ & 2 & 1 & 0 \\ \hline$\Z_{8-II}$ & $\lp\bar{z_2},\bar{z_1},\bar{z_3}\rp$ & 2 & 1 & 1 \\ \hline\multirow{2}{*}{$\Z_{12-I}$} & $\lp\bar{z_3},\bar{z_2},\bar{z_1}\rp$ & 2 & 1 & 0 \\ & $\lp\bar{z_3},\bar{z_2},i\bar{z_1}\rp$ & 2 & 1 & 0 \\ \hline\multirow{2}{*}{$\Z_{12-II}$} & $\lp\bar{z_2},\bar{z_1},\bar{z_3}\rp$ & 2 & 1 & 1 \\ & $\lp\bar{z_2},e^{\pi i/3}\bar{z_1},\bar{z_3}\rp$ & 2 & 1 & 1 \\ \hline$\Z_2\times\Z_2$ & $\lp\bar{z_1},\bar{z_3},\bar{z_2}\rp$ & 2 & 1 & 3 \\ \hline\multirow{2}{*}{$\Z_2\times\Z_4$} & $\lp\bar{z_1},\bar{z_3},\bar{z_2}\rp$ & 2 & 1 & 1 \\ & $\lp\bar{z_1},\bar{z_3},i\bar{z_2}\rp$ & 2 & 1 & 1 \\ \hline$\Z_2\times\Z_6$ & $\lp\bar{z_1},\bar{z_3},\bar{z_2}\rp$ & 2 & 1 & 1 \\ \hline\multirow{2}{*}{$\Z_2\times\Z_6'$} & $\lp\bar{z_1},\bar{z_3},\bar{z_2}\rp$ & 2 & 1 & 0 \\ & $\lp\bar{z_2},\bar{z_1},\bar{z_3}\rp$ & 2 & 1 & 0 \\ \hline$\Z_3\times\Z_3$ & $\lp\bar{z_1},\bar{z_3},\bar{z_2}\rp$ & 2 & 1 & 0 \\ \hline\multirow{2}{*}{$\Z_3\times\Z_6$} & $\lp\bar{z_1},\bar{z_3},\bar{z_2}\rp$ & 2 & 1 & 0 \\ & $\lp\bar{z_1},\bar{z_3},-\bar{z_2}\rp$ & 2 & 1 & 0 \\ \hline$\Z_4\times\Z_4$ & $\lp\bar{z_1},\bar{z_3},\bar{z_2}\rp$ & 2 & 1 & 0 \\ \hline$\Z_6\times\Z_6$ & $\lp\bar{z_1},\bar{z_3},\bar{z_2}\rp$ & 2 & 1 & 0 \\ \hline\end{tabular}
\label{table:nonstandardorientifolds}
\end{table}

\subsection{Turning on NSNS fluxes}\label{subsec:fluxes}

Now for each of the orientifolds discussed in the last subsection, we would like to understand which generalized fluxes can be turned on consistently.  The discussion will be very brief, and we will refer the interested reader to \cite{Ihl:2007ah}, \cite{Robbins:2007yv} for a more careful discussion of our approach.

As explained in section \ref{subsec:generalized}, the $H$-flux and metric fluxes we would like to turn on can be thought of in terms of their components $H_{ijk}$ and $f^i_{jk}$, where the indices run over the six legs of the torus.  For each toroidal orientifold, we need these objects to transform correctly under the quotient group.  This means that both $H_{ijk}$ and $f^i_{jk}$ must be invariant under the orbifold group, and $H_{ijk}$ should be odd under the action of $\s$, while $f^i_{jk}$ should be even.

Next, we must also impose the Bianchi identities, which in this case take the form
\be
H_{i[jk}f^i_{\ell m]}=0,\qquad f^i_{j[k}f^j_{\ell m]}=0.
\ee
In terms of the geometry of the underlying twisted torus, the latter equation is simply that the exterior derivative is nilpotent ($d^2=0$), and the former condition is simply that $H$ is a closed three-form ($dH=0$).  In section \ref{sec:Rulingoutmodels} we will tabulate the fluxes that can be turned on and all solutions to the Bianchi identities for each of our models.  In principle one could violate the Bianchi identities by including localized NSNS sources \cite{Villadoro:2007tb}, but we will not include such objects in this paper.

Once we have found the set of independent fluxes which survive the orientifold quotient, it turns out to be more convenient, for the purposes of the low-energy theory, to refer not to the components $H_{ijk}$ and $f^i_{jk}$, but instead to certain coefficients $p_K$, $r_{aK}$, and $\hat{r}_\al^K$, given by
\be
H=p_Kb^K,\qquad d\om_a=-r_{aK}b^K,\qquad d\mu_\al=-\hat{r}_\al^Ka_K,
\ee
where $a_K$, $b^K$, $\om_a$, and $\mu_\al$ are forms which descend from the untwisted cohomology of the torus {\it without} fluxes.  There is an invertible linear map between the $p_K$ and the independent components $H_{ijk}$.  Similarly, the coefficients $r_{aK}$ and $\hat{r}_\al^K$ are always given by linear combinations of the independent $f^i_{jk}$, but in this case the map is not always invertible; there can be more independent $f^i_{jk}$ than $r_{aK}$ and $\hat{r}_\al^K$.  In these cases, the extra metric fluxes do appear in the Bianchi identities, and should be taken into account when classifying the solutions, but the scalar potential and the tadpole constraints (see below) depend only on $r_{aK}$ and $\hat{r}_\al^K$.

The Ramond-Ramond tadpoles are given by\footnote{This expression is actually not exactly correct.  Rather, this is a cohomological condition (in the sense of the cohomology of the torus without metric fluxes).  There is an exact tadpole constraint of the schematic form $DF=J$, where $D$ is the generalized derivative ($d+H$ on the twisted torus), $F$ is the formal sum of the RR fluxes, and $J$ is a delta function form describing the configuration of O6-planes and D6-branes.}
\be
-\sqrt{2}\lp p_K\widetilde{m}-r_{aK}m^a\rp=2N_K^{(\mathrm{O6})}-N_K^{(\mathrm{D6})},
\ee
where the right hand side of the equation above represents the contribution from localized sources, both O6-planes, which sit at the fixed points of the orientation reversing elements of the orientifold group, and the D6-branes which we allow to be added anywhere.  We will not really be viewing these tadpoles as constraints in the present work, taking the attitude that D6-branes can be added as needed.  Clearly, for a detailed analysis of any given model, one would have to proceed more carefully, taking into account RR quantization as well as the action of the orientifold on the open string sector.

From the perspective of the low-energy effective theory, these seem to be the only constraints that need to be obeyed.  Indeed, we will find that for most models, these constraints are already enough for us to be able to apply our no-go theorems and rule out slow-roll inflation and de Sitter extrema from the corresponding scalar potential.  However, we would like to understand exactly which models can be constructed consistently from a ten-dimensional perspective.  There are at least two approaches to this problem, via the coset constructions of \cite{Tomasiello:2007eq}, \cite{Koerber:2008rx}, \cite{Caviezel:2008ik}, and the base-fiber twisted torus constructions of \cite{Ihl:2007ah}, which are inspired by \cite{Hellerman:2002ax}, \cite{Dabholkar:2002sy} and others.  In the present work we will focus mainly on the latter approach when we want explicit constructions, so we will now briefly review it.

For a given orientifold of $T^6$, we first pick a division of the torus into a base and a fiber, in such a way so that the orientifold group does not mix the two.  More precisely, we require the tangent spaces of the base and fiber to be invariant subspaces of the orientifold action on the tangent space of the $T^6$.  Once this splitting has been chosen, then for each direction in the base, labeled by an index $i=1,\ldots,n$, we will choose a matrix $M_i\in\mathfrak{so}(6-n,6-n)$.  The entries of these matrices will correspond to the components of our fluxes,
\be
\lp M_i\rp=\lp\begin{matrix}-f^b_{ia} & H_{iab} \\ -Q^{ab}_i & f^a_{ib}\end{matrix}\rp.
\ee
Here $a$ and $b$ are indices in the fiber directions.  $Q^{ab}_i$ are non-geometric fluxes which we don't want to turn on in this work, so the matrices in our examples will be upper-block-diagonal.  Note that using these constructions we can only turn on sets of fluxes that have one lower index lying along the base, and the other two indices (upper or lower) lying along the fiber.

Using this language, it is straightforward to restrict to matrices that are compatible with the orientifold group.  The Bianchi identities are reproduced simply by demanding that $M_i$ and $M_j$ commute for all $i$ and $j$.  Finally, quantization conditions for the generalized fluxes become simply the condition
\be
\exp\ls\vec{\lambda}\cdot\vec{M}\rs\in\SO(6,6;\Z),\qquad\vec{\lambda}\in\Lambda\cong\Z^6
\ee
where $\Lambda\cong\Z^6$ is the lattice of torus identifications embedded in the tangent space of $T^6$, and where we identify the $M_i$ with a Lie algebra valued cotangent vector.  One of the great values of the base-fiber construction is that it enables one to identify the correct quantization conditions on the generalized fluxes, something that is not apparent from considerations of the low energy theory alone.  And, in fact, there can be cases where the quantization conditions forbid us from turning on certain components of metric fluxes, for instance.  These constructions are slightly generalized and formulated more precisely in \cite{Bergman:2007qq}.

\subsection{Residual symmetries}\label{subsec:ressym}

Each of the no-go theorems we derived requires certain assumptions about which fluxes can be non-vanishing.  Sometimes a solution of the Bianchi identities will automatically satisfy the assumptions for one of our theorems, but there will also be cases which do not appear to fall into one of these cases, but which we found numerically to still satisfy a bound on $\e$.  In almost all of these cases we were able to find a symmetry (that is a field redefinition that preserved the form of the potential while simply changing which fluxes were turned on) that mapped us into a configuration for which the no-go theorems apply.  For this reason, it is important to identify the group of symmetries which can act in this way.

This turns out to be fairly straightforward.  Recall that type IIA on $T^6$ has a group of T-duality symmetries $\SO(6,6;\Z)$, or, somewhat more precisely, we should use the double cover, $\operatorname{Spin}(6,6;\Z)$.  This group includes large diffeomorphisms of the torus (living in a $\GL(6;\Z)$ subgroup), shifts of the $B$-field, and also non-geometric symmetries such as performing a T-duality on a $T^2\subset T^6$.  In fact, the orientifold group $\widehat{G}$ by which we are to quotient can also be considered as a subgroup of this T-duality group.  The orbifold group $G$ sits inside $\SU(3)\subset\GL(6;\Z)$, while the orientation-reversing elements of $\widehat{G}$ sit inside of a $\Z_2$ extension of $\GL(6;\Z)$ in $\operatorname{Spin}(6,6;\Z)$ \cite{Ihl:2007ah}.  The resulting space will still have a group of duality symmetries given by the elements $h$ of the full T-duality group which satisfy
\be
h\widehat{G}h^{-1}=\widehat{G}.
\ee

All of the fields and fluxes (here we should include the full set of non-geometric fluxes), as well as the Bianchi identities and tadpole constraints, will transform as representations of these residual symmetries.  We will find situations where we can use these symmetries to map one set of fluxes which solves the Bianchi identities, to a new set of fluxes which still solves the Bianchi identities, but also satisfy the assumptions of one of our no-go theorems.  The resulting bound on $\e$ will apply to both configurations of fluxes (since we have simply performed a field redefinition).

\section{Application of the No-Go Theorems to Toroidal Orientifolds}
\label{sec:Rulingoutmodels}

In this section we will apply our no-go theorems to the toroidal orientifold models discussed above. Since we are restricting to the untwisted sector we have eleven different models that are uniquely determined by their triple intersection numbers \eqref{eq:tripleintersect} and the prepotential in the complex structure sector \eqref{eq:prepot}. These models are summarized in table \ref{table:models}.
\begin{table}[h!]
\begin{tabular}{|c|c|c|c|c|c|c|}
  \hline
  $\sharp$&$\! \! h^{2,1} \! \!$ & $\! \! h^{1,1}_+\! \!$ & $\! \!h^{1,1}_-\! \!$ & $\kappa,\, \hat{\kappa}$ & $p_n(\mathcal{Z})\,(=1)$ & Quotients \\ \hline \hline
  I &0 & 0 & 3 & $\kappa_{123}=1$ & $2 \mathcal{Z}^1$ & $\begin{array}{c}
                                                                 \mathbb{Z}_7, \mathbb{Z}_{8-I},\mathbb{Z}_{12-I}, \\
                                                                 \!\! \mathbb{Z}_2 \times \mathbb{Z}_{6'}, \mathbb{Z}_3 \times \mathbb{Z}_3,\!\!\\
                                                                 \!\! \mathbb{Z}_3 \times \mathbb{Z}_6, \mathbb{Z}_4 \times \mathbb{Z}_4, \!\!\\
                                                                  \mathbb{Z}_6 \times \mathbb{Z}_6
                                                               \end{array}$\\ \hline
  II &0 & 1 & 2 & $\kappa_{122}=1,\, \hat{\kappa}_{111}=-1$ & $2 \mathcal{Z}^1$ & $\begin{array}{c}
                                                                 \mathbb{Z}_{8-I}, \, \mathbb{Z}_{12-I},\\
                                                                 \!\! \mathbb{Z}_2 \times \mathbb{Z}_{6'}, \mathbb{Z}_3 \times \mathbb{Z}_3, \!\!\\
                                                                 \!\!\mathbb{Z}_3 \times \mathbb{Z}_6, \mathbb{Z}_4 \times \mathbb{Z}_4,\!\! \\
                                                                 \mathbb{Z}_6 \times \mathbb{Z}_6
                                                               \end{array}$\\ \hline
  III &0 & 1 & 4 & $\begin{array}{c}
                                                                \kappa_{123}=1,\kappa_{144}=-1, \\
                                                                 \hat{\kappa}_{111}=-1
                                                               \end{array}$ & $2 \mathcal{Z}^1$ & $\mathbb{Z}_{6-I}$\\ \hline
  IV &0 & 2 & 3 & $\begin{array}{c}
                                                                \kappa_{122}=1,\hat{\kappa}_{1\alpha\alpha}=-1, \\
                                                                 \alpha \in \{1,2,3\}
                                                               \end{array}$ & $2 \mathcal{Z}^1$ & $\mathbb{Z}_{6-I}$\\ \hline
  V &0 & 3 & 6 & $\begin{array}{c}
                                                                 \kappa_{123}=1,\kappa_{456}=-2,\\
                                                                  \!\! \!\! \! \! \!\! \!\! \! \kappa_{144} = \kappa_{255} = \kappa_{366}=-2,\! \!\! \!\! \! \! \!\! \! \!\\
                                                                 \! \!\! \! \! \!\! \! \! \! \!\hat{\kappa}_{111} = \hat{\kappa}_{222} = \hat{\kappa}_{333}=-1,\! \! \!\! \!\! \! \! \!\! \!\\
                                                                 \! \!\hat{\kappa}_{423} = \hat{\kappa}_{513}= \hat{\kappa}_{612}=1,\! \!\\
                                                               \end{array}$ & $2 \mathcal{Z}^1$ & $\mathbb{Z}_{3}$ \\ \hline
  VI & 1 & 0 & 3 & $\kappa_{123}=1$ & $2^2 \mathcal{Z}^1 \mathcal{Z}^2 $ & $\begin{array}{c}
                                                                \! \!  \!\mathbb{Z}_{6-II}, \mathbb{Z}_{8-II}, \mathbb{Z}_{12-II},\! \! \! \\
                                                               \!\!\mathbb{Z}_2 \times \mathbb{Z}_4, \mathbb{Z}_2 \times \mathbb{Z}_6\!\!
                                                               \end{array}$ \\ \hline
  VII & 1 & 1 & 2 & $\kappa_{122}=1, \hat{\kappa}_{111}=-1$ & $2^2 \mathcal{Z}^1 \mathcal{Z}^2 $& $\begin{array}{c} \mathbb{Z}_{8-II}, \, \mathbb{Z}_{12-II},\\
  \!\!\mathbb{Z}_{2} \times \mathbb{Z}_{4}, \mathbb{Z}_{2} \times \mathbb{Z}_{6}\!\! \end{array}$\\ \hline
  VIII & 1 & 1 & 4 & $\begin{array}{c}
                                                                 \kappa_{123}=1,\kappa_{144}=-1, \\
                                                                 \hat{\kappa}_{111}=-1
                                                               \end{array}$ & $2^2 \mathcal{Z}^1 \mathcal{Z}^2$ & $\mathbb{Z}_{4}$\\ \hline
  IX & 1 & 3 & 2 & $\begin{array}{c} \kappa_{122}=1, \hat{\kappa}_{1\alpha \alpha}=-1,\\ \alpha \in \{1,2,3\} \end{array}$ & $2^2 \mathcal{Z}^1 \mathcal{Z}^2 $& $\mathbb{Z}_{4}$ \\ \hline
  X & 3 & 0 & 3 & $\kappa_{123}=1$ & $2^4 \mathcal{Z}^1 \mathcal{Z}^2 \mathcal{Z}^3 \mathcal{Z}^4$& $\mathbb{Z}_{2} \times \mathbb{Z}_{2}$ \\ \hline
  XI & 3 & 1 & 2 & $\kappa_{122}=1, \hat{\kappa}_{111}=-1$ & $\!  2^4 (\mathcal{Z}^1)^2 (\mathcal{Z}^2 \mathcal{Z}^3 -(\mathcal{Z}^4)^2)\!$ & $\mathbb{Z}_{2} \times \mathbb{Z}_{2}$ \\ \hline
\end{tabular}
\caption{This table summarizes the models we are considering. It contains the number of invariant forms and the non-vanishing triple intersection numbers \eqref{eq:tripleintersect} together with the prepotential for the complex structure sector \eqref{eq:prepot}.}\label{table:models}
\end{table}
Note that for all but the $\mathbb{Z}_3$ quotient we have a factorization in the K\"ahler sector and for all models with $h^{2,1}>0$ we have a factorization in the complex structure sector. 

For the benefit of the reader primarily interested in the results rather than how they arise, we summarize the weakest bounds on the slow-roll parameter $\epsilon$ for the various cases in table~\ref{tab:summary}. 
\begin{table}[h!]
\centering
\begin{tabular}{|c|c|c|c|c|}\hline
Case & I,II,III,IV,V & VII & I',II', VI, VIII, IX & X, XI \\\hline
$\epsilon\geq$ &$\frac{27}{13}$ & 2 & $\frac95$ &0 \\\hline
\end{tabular}
\caption{This table presents a summary of the weakest bounds on the slow-roll parameter $\epsilon$ for the various cases. The $Z_{8-I}$-quotient turns out to be special for both case I and II, and we denote it by I' and II'.}
\label{tab:summary}
\end{table}
There are several special cases that can be shown to satisfy stronger bounds. These are omitted from table \ref{tab:summary}, but they are discussed in some detail below.

We will now discuss the solutions to the Bianchi identities for all of these cases and check which of our no-go theorems can be applied. We have also minimized $\e$ numerically by allowing the moduli and fluxes to vary. We found generically that the bound given by the no-go theorem can be attained which proves that it is impossible to derive a stronger no-go theorem. For the two models X and XI we will find solutions to the Bianchi identities that escape all of our no-go theorems. In these cases it is possible to get vanishing $\epsilon$ and we will discuss this in detail in the next section.

\subsection{Case I}
There are two special quotients $\mathbb{Z}_7$ and $\mathbb{Z}_{8-I}$ in this first case. Both of these have extra metric fluxes that are not contained in the matrix $r_{aK}$.\\
The generic solution to the Bianchi identities for all models in case I has only one of the three entries in the $r$ vector non-zero. (For $\mathbb{Z}_7$ and $\mathbb{Z}_{8-I}$ the extra metric fluxes are zero.) For these solutions there is an $\SL(2,\R)$ subgroup of the residual T-duality symmetries (see section \ref{subsec:ressym}) under which $p_1$ and the nonvanishing $r$-flux transform as a doublet.  This symmetry can be used to set the $r$-flux to zero (note that if other components of $r$ were nonzero, then these T-dualities would mix them with nongeometric fluxes). A more pedestrian way to see this is that a shift in one of the $B$-axions allows us to set $p_1=0$. A T-duality then takes us to a configuration that has no metric flux, and we find the bound $\e \geq \frac{27}{13}$. \\
For $\mathbb{Z}_{8-I}$ there is one more solution to the Bianchi identities due to the extra metric fluxes, which we call $f_1$ and $f_2$. It reads $r_{21} =0,\, r_{11} r_{31} = -f_1^2 = - f_2^2$. The factorization in the K\"ahler sector allows us to apply our no-go theorem since $r_{21} = "r_{01}"=0$, and we obtain $\e \geq \frac95$.

\subsection{Case II}
As before, $\mathbb{Z}_{8-I}$ is special because it has two extra metric fluxes $f_1, f_2$ that are not contained in the $r$ matrix.\\
The solution to the Bianchi identities common to all quotients forces $r_{21} = \hat{r}_1^1=0,\, (f_1=f_2=0)$ so that we are again left with only one single metric flux that can be mapped to $H$-flux just as discussed in case I, and we again find $\e \geq \frac{27}{13}$.\\
For $\mathbb{Z}_{8-I}$ there is one more solution to the Bianchi identities $r_{11}=r_{21}=p_1=0$, $\hat{r}_1^1= f_1^2 +f_2^2$. \\
From the factorization in the K\"ahler sector and using $r_{11} = "r_{0K}"=0$ we find $\e \geq \frac95$.

\subsection{Case III}
This case has two solutions to the Bianchi identities $r_{21} =r_{31} =r_{41} = \hat{r}_1^1=0$ and $r_{11} =2 r_{21} r_{31} -r_{41}^2 = \hat{r}_1^1 = 0$. The second case can be brought into a form where only one of the $r_{21}, \, r_{31}, \, r_{41}$ is non-zero using field redefinitions as described in subsection \ref{subsec:ressym}. To see this, note that the potential has an $SO(2,1)$ symmetry under which $r_{21}$, $r_{31}$, and $r_{41}$ transform as the components of a null vector. This means we can boost to a configuration that has only $r_{21}$ or $r_{31}$ non-zero. After shifting one of the $B$-axions to absorb $p_1$, both configurations can be T-dualized to a case with only $H$-flux resulting in $\e \geq \frac{27}{13}$.

\subsection{Case IV}
This case again has only one non-vanishing entry in the $r$ matrix since the Bianchi identities are $r_{21} = \hat{r}_\alpha^1=0, \, \forall \alpha \in \{ 1,2,3\}$. Using a shift in one of the axions together with a T-duality transformation, we see that the configuration is equivalent to one with $H$-flux and no metric flux. Therefore, we find $\e \geq \frac{27}{13}$.

\subsection{Case V}
The Bianchi identities are $\hat{r}_\alpha^1=0$, $2 r_{11} r_{41}+r_{51} r_{61} =2 r_{21} r_{51}+r_{41} r_{61}=2 r_{31} r_{61}+r_{41} r_{51}=0, 4 r_{11} r_{21} -r_{61}^2 = 4 r_{11} r_{31} - r_{51}^2=4 r_{21} r_{31} - r_{41}^2=0$.    These have two classes of solutions.

The first class is characterized by the vanishing of two of the fluxes $r_{41}$, $r_{51}$, and $r_{61}$. With the others being related to this by symmetries, let us consider $r_{51}=r_{61}=0$, for definiteness. The remaining Bianchi identities then are $r_{11} r_{21}=0$, $r_{11} r_{31}=0$, and $4r_{21} r_{31}-r_{41}^2=0$. One obvious solution is $r_{21}=r_{31}=r_{41}=0$, but this is equivalent by symmetries to a special case of the solution $r_{11}=0$, and $4r_{21} r_{31} -r_{41}^2=0$, so let us focus on the latter solution. The potential has a manifest $SO(2,1)$ symmetry under which $r_{21}, r_{31}, r_{41}$ transform as the components of a vector. The Bianchi identity enforces this vector to be null. We can perform a boost such that only $r_{21}$ or $r_{31}$ is non-zero. At least locally, we can also redefine one of the $B$-axions to set $p_1=0$. It is then easy to see that the remaining configuration is T-dual to a configuration with H-flux but no metric fluxes, which implies $\epsilon\geq\frac{27}{13}$.

A naively inequivalent class of solutions has $r_{41}$, $r_{51}$, and $r_{61}$ non-zero. The Bianchi identities then determine $r_{11}$, $r_{21}$, $r_{31}$ in terms of these as $r_{11}=-r_{51} r_{61}/2r_{41}$, $r_{21}=-r_{41} r_{61}/2r_{51}$, $r_{31}=-r_{51} r_{61}/2r_{41}$. This class turns out to be related by symmetries to the first class, and consequently also obeys $\epsilon\geq\frac{27}{13}$.  To see this, we can pick our favorite among $r_{11}$, $r_{21}$, and $r_{31}$, which, without loss of generality, we will take to be $r_{21}$. Let us make use of another $SO(2,1)$ symmetry of the potential under which $r_{21}$  transforms as a scalar, $r_{11}$, $r_{31}$, and $r_{51}$ as the components of a vector, and $r_{41}$ and $r_{61}$ as the components of a spinor. We can use the $SO(2)$ subgroup of $SO(2,1)$ to set one component of the spinor to zero, say, $r_{61}$. The Bianchi identities together with the invariance of $r_{21}$ guarantee that as we take $r_{61}$ to zero, because $r_{41}$ remains finite, $r_{51}$ must vanish such that $r_{41} r_{61}/r_{51}$ remains constant. With $r_{51}$ and $r_{61}$ zero, we are back to the first class of solutions, and a boost of our earlier $SO(2,1)$ group followed by a field redefinition of one of the axions and a T-duality again takes us to a configuration without metric flux implying $\epsilon\geq\frac{27}{13}$.

\subsection{Case VI}
The $\mathbb{Z}_{6-II}$ quotient is special because it allows for one extra metric flux we denote $f_1$ that is not contained in the $r$ matrix.\\
We can exclude all solutions by using our two no-go theorems that rely on the factorization in the K\"ahler sector. There are two solutions up to permutation of the $a \in  \{1,2,3\}$ index, that are common to all models (and for which the extra flux $f_1$ vanishes). The first has $r_{2K} =r_{3K} =0 ,\, \forall K \in  \{1,2 \}$ and our no-go theorem gives $\e \geq 2$. The other solution reads $r_{1K} = r_{21} r_{32} + r_{22} r_{31} =0, \, \forall K  \in \{1,2\}$ and we find  $\e \geq \frac95$.\\
Finally, for $\mathbb{Z}_{6-II}$ we have one more solution for which the extra metric flux is non-vanishing: $r_{1K} =0, \, \forall K \in \{1,2\},$ $r_{2L} = r_{3M} = 0,$ $r_{2M} r_{3L} = f_1^2$, $L \neq M,$ $L, M  \in \{1,2\}$. We again find $\e \geq \frac95$.

\subsection{Case VII}
Here we have four different solutions to the Bianchi identities that can be dealt with using four different no-go theorems. \\
The first solution is $\hat{r}^K = r_{1K} =0,\, \forall K \in \{1,2 \}$ and $r_{21}$ or $r_{22} =0$. We are left with one single entry in the $r$ matrix. A field redefinition relates this to a configuration with only $H$-flux and we again have $\e \geq \frac{27}{13}$.\\
Another solution is $\hat{r}^K = r_{2K} = 0,\, \forall K \in \{1,2 \}$ and our no-go theorem for factorization in the K\"ahler sector gives $\e \geq 2$.\\
The third solution is $r_{2K}=0,\, \forall K \in \{1,2 \}$, $\hat{r}^L = p_M =r_{1M}=0,$ $L \neq M,$ $L, M \in \{1,2 \}$. Here we can use the factorization in $p_n(\mathcal{Z})$ and apply one of our no-go theorem for $n_1 = n_2=1$ and find $\e \geq \frac{4 n_2}{n_1 + n_2} =2$.\\
Using the factorization in both complex and K\"ahler sector we can show that the last solution $r_{1K}=0,\, \forall K \in \{1,2 \}$, $\hat{r}^L = p_M =r_{2M}=0,$ $L \neq M,$ $L, M \in \{1,2 \}$ has $\e \geq \frac{9 n_1 + 5 n_2}{5n_1+n_2} = \frac73$, where we used $n_1=n_2=1$.

\subsection{Case VIII}
This case has six different solutions to the Bianchi identities. Three of those are
\begin{itemize}
\item $r_{1K}=\hat{r}_1^K=0, \, \forall K \in \{1,2 \}, \, r_{21} r_{32} +r_{22} r_{31}- r_{41}r_{42}=0$

\item $r_{1K}= r_{4K}=p_K \hat{r}_1^K=r_{2K} \hat{r}_1^K=r_{3K} \hat{r}_1^K=0, \, \forall K \in \{1,2 \}, \,2 r_{2L} r_{3L}+ (\hat{r}_1^M)^2 = \epsilon^{LM} p_L r_{2M} =0, 2 r_{2L} r_{3M}-\hat{r}_1^1 \hat{r}_1^2 =0, \epsilon^{LM} p_L r_{3M}=0, L \neq M, \,L,M \in \{1,2\}$

\item $r_{1K}=p_K \hat{r}_1^K=r_{aK} \hat{r}_1^K=0, \, \forall K \in \{1,2 \}, \, 2 r_{2L} r_{3L} - (r_{4L})^2 + (\hat{r}_1^M)^2=0, \,\epsilon^{LM} p_L r_{aM} =0, \, 2 r_{2L} r_{3M} - r_{41} r_{42} - \hat{r}_1^1 \hat{r}_1^2=0, \, \epsilon^{LM} r_{4L} r_{aM} =0, L \neq M, \,L,M \in \{1,2\}$
\end{itemize}
These all have $r_{1K}=0$ and we can use the factorization in the K\"ahler sector to show that $\e \geq \frac95$.\\
The next solution $\hat{r}_1^K=r_{2K}=r_{3K}=r_{4K}=0 , \, \forall K \in \{1,2 \}$ has $\e \geq 2$ again due to the factorization in the K\"ahler sector.\\
The fifth solution $r_{2K}=r_{3K}=r_{4K}=0,\forall K \in \{1,2 \},$ $\hat{r}_1^L=p_M=r_{1M}=0$, $L\neq M,\, L,M \in \{1,2\}$ has $\e \geq \frac{4 n_2}{n_1 + n_2} =2$. This follows from the no-go theorem that relies on the factorization of the complex structure sector and we have used $n_1 = n_2=1$.\\
The last case $r_{1K}=0, \, \forall K \in \{1,2 \}$ $\hat{r}_1^L=p_M=r_{2M}= r_{3M}=r_{4M}=0, \, L\neq M,\, L,M \in \{1,2\} $, can be dealt with using the no-go theorem that relies on the factorization in both complex and K\"ahler sector and has $\e \geq \frac{9 n_1 + 5 n_2}{5n_1+n_2} = \frac73$ since $n_1=n_2=1$.

\subsection{Case IX}
In this case we again need several different no-go theorems.\\
The first class of solutions $r_{1K} =\hat{r}^K_\alpha=0, \forall K \in \{1,2 \}, \, \forall \alpha \in \{1,2,3\}$ and $r_{21} =0$ or $r_{22}=0$ has only one non-vanishing metric flux, and the configuration is dual to one with only $H$-flux, so we find $\e \geq \frac{27}{13}$.\\
The next solution $r_{2K} =\hat{r}^K_\alpha=0, \forall K \in \{1,2 \}, \, \forall \alpha \in \{1,2,3\}$ gives $\e \geq 2$ due to the factorization in the K\"ahler sector.\\
From the factorization in the complex structure sector we find $\e \geq \frac{4 n_2}{n_1 + n_2} =2$ for $n_1=n_2=1$ for the third solution $r_{2K}=0, \, \forall K\in\{1,2\}$ and $\hat{r}_1^L = p_M=r_{2M} =0,\, L\neq M,\, L,M \in \{1,2\}$.\\
The next solution $r_{aK}=p_K = \sum_\alpha \hat{r}_\alpha^1 \hat{r}_\alpha^2 =0, \, \forall a,K \in \{1,2\}$ satisfies the condition for our no-go based on the factorization in the K\"ahler sector and we find $\e \geq \frac95$.\\
Using the factorization in both complex and K\"ahler sector we can show that the last solution $r_{1K}=0,\, \forall K \in \{1,2 \}$, $\hat{r}_\alpha^L = p_M =r_{2M}=0, \, \forall \alpha \in \{1,2,3\},$ $L \neq M,$ $L, M \in \{1,2 \}$ has $\e \geq \frac{9 n_1 + 5 n_2}{5n_1+n_2} = \frac73$, where we used $n_1=n_2=1$.

\subsection{Case X}
Here the solutions to the Bianchi identities can be grouped into five different classes. \\
The first class of solutions has two of the three rows in the $r$ matrix equal to zero. The third one is arbitrary and can be identified with $"r_{0K}"$ and our no-go theorem that relies on the factorization of the K\"ahler sector can be used to obtain $\e \geq 2$.\\
The next case has partially non-vanishing entries in two rows and at least two columns. The entire third row is zero so that we have this time $"r_{0K}=0"$ and $\e \geq \frac95$.\\
The third class encompasses four solutions that each have only three non-vanishing entries with one in each row. The non-vanishing $r$ components for the four different cases are $1) \, r_{11}, r_{24}, r_{33}\neq0$, $\,2) \, r_{12}, r_{23}, r_{34}\neq0$, $\, 3) \, r_{13}, r_{22}, r_{31}\neq 0$, $\, 4) \, r_{14}, r_{21}, r_{32}\neq0$. Each of these cases leads numerically to $\e \approx 1.57721$. We leave it up to the interested reader to try to find a no-go theorem that gives this value using the residual symmetries and factorization in both K\"ahler and complex structure sector.\\
The fourth class of solutions has one of the four columns of the $r$ matrix non-vanishing and arbitrary. Numerically one obtains $\e \approx \frac43$ and we leave it again to the interested reader to find the corresponding no-go theorem. \\
The last class has two, three, or all four columns non-zero. The non-zero metric fluxes are not all independent but have to satisfy constraints that result from the Bianchi identities. The most generic case has all twelve entries in the $r$ matrix non-zero. There are six constraints so that we are left with six independent metric fluxes. For this class there cannot be a bound on $\e$ from a no-go theorem since one can find numerically solutions that have $\e \approx 0$. We will discuss an explicit example in more detail in subsection \ref{standardZ2}.

\subsection{Case XI}
The solutions to the Bianchi identities can be again grouped into five classes.\\
The first class has $r_{11} = r_{21} = p_1 = \hat{r}_1^2= \hat{r}_1^3= \hat{r}_1^4 =0$ and $2 r_{13} r_{22}+2 r_{12} r_{23} -r_{14} r_{24}=0$. From the factorization in the complex sector we find $\e \geq \frac{4 n_2}{n_1 + n_2} =2$, where now $n_1= n_2 =2$.\\
The next class has $r_{1K} =0, \forall K \in \{1,2,3,4\}$. The remaining NSNS fluxes $r_{2K}, \hat{r}_1^K, p_K$ are constrained by the remaining Bianchi identities. We can use the factorization in the K\"ahler moduli sector since $r_{1K} ="r_{0K}" =0$ to obtain $\e \geq \frac95$.\\
The third class has $r_{21} = \hat{r}_1^1 =0$ and $r_{11} \neq 0$. The Bianchi identities then reduce to $p_K \hat{r}_1^K = r_{1K} \hat{r}_1^K =0$ and $r_{22} = \frac{-r_{14} \hat{r}_1^3 -2 r_{12} \hat{r}_1^4}{2 r_{11}},$ $r_{23} = \frac{r_{14} \hat{r}_1^2 +2 r_{13} \hat{r}_1^4}{2 r_{11}},$ $r_{24} = \frac{r_{12} \hat{r}_1^2 - r_{13} \hat{r}_1^3}{r_{11}}$. None of the no-go theorems apply and numerically one finds $\e \approx 0$. We will discuss this class and the next two that all allow for extremal points with very small $\e$ in subsection \ref{nonstandardZ2} in greater detail.\\
The next class has $r_{11}=r_{21} = \hat{r}_1^1=0$. The remaining fluxes are again constrained by the Bianchi identities. Numerically we find vanishing $\e$ in the limit where we have $\hat{r}_1^K \approx 0, \forall K \in \{1,2,3,4\}$. In this limit the only Bianchi identity that constrains the non-vanishing NSNS fluxes is $2 r_{13} r_{22}+2 r_{12} r_{23} -r_{14} r_{24}=0$. \\
The last class of solutions has $r_{11}=0$ but $r_{21} \neq 0$. Again there are several Bianchi identities that constrain the remaining fluxes. Nevertheless, it is generically possible to obtain small $\e$ so that there cannot be a no-go theorem. We will present some of the details of our numerical studies of this case in the next section.

\section{Examples with Small $\epsilon$}
\label{sec:smalleexamples}
In the previous section, we have shown that our no-go theorems rule out slow-roll inflation and de Sitter vacua in large classes of models. However, there were solutions to the Bianchi identities for the two $\mathbb{Z}_{2} \times \mathbb{Z}_{2}$ orbifold models that escaped all no-go theorems. In this section we give explicit examples for the models that have regions in moduli space with (very) small $\e$ that likely correspond to de Sitter extrema.\footnote{Since the analysis is purely numerical, it is impossible to tell whether these are extrema with all derivatives of the potential vanishing. All we know is that the values we find are compatible with zero to within our working precision, but they might just be very shallow.}
Since we are now presenting explicit solutions rather than no-go theorems some words of caution are in order. ``Adding metric fluxes'' to an existing geometry only leads to a well-defined compact space if the fluxes are properly quantized. One way to explicitly construct a space with metric fluxes is the base-fiber splitting framework~\cite{Hellerman:2002ax}, \cite{Dabholkar:2002sy}, \cite{Ihl:2007ah}. This framework also allows to derive the quantization conditions for all the NSNS fluxes. We will therefore check whether we can explicitly construct the compact spaces that lead to de Sitter extrema or whether the base-fiber approach is not compatible with the solutions to the Bianchi identities that gave small $\e$. Further constraints on these models arise from the quantization of the RR fluxes and the tadpole cancellation condition. To ensure the validity of the supergravity approximation, one also has to check that the volume of the internal space is large in string units and that the string coupling is small. Since we always find at least one tachyonic direction for the extremal points with vanishing $\e$ we will only consider the restrictions from the base-fiber constructions in the subsections below.

\subsection{Standard orientifold of $\Z_2\times\Z_2$}\label{standardZ2}

The solutions to the Bianchi identities that allow for vanishing $\e$ have at least non-vanishing entries in two columns and all three rows. For simplicity we take the case where $r_{a3}= r_{a4}=0,\, \forall\, a \in \{1,2,3\}$. Assuming that $r_{22} \neq 0 $ we can solve the remaining Bianchi identities and find $r_{11} = -\frac{r_{12} r_{21}}{r_{22}}, \, r_{31} = \frac{r_{21} r_{32}}{r_{22}}$. Minimizing $\e$ by letting all the moduli and remaining fluxes vary, we find $\e \approx 10^{-21}$ for the following values\footnote{For cosmetic reasons, the following values are rounded to four digits and give $\e \approx 10^{-4}$.} of the fluxes and moduli\footnote{The $C_3$ axions $\xi^K$ appear in the potential only through the linear combinations $p_K \xi^K$ and $r_{aK} \xi^K$. Since in this simple case $r$ has rank two we can stabilize only three linear combinations of them. In particular $\xi^3-\xi^4$ is a flat direction. By allowing for at least three non-vanishing columns in the $r$ matrix one finds examples without flat directions.}
\begin{align*}
\widetilde{m} &\approx -.2026, \, m_1 = m_2=m_3 \approx .6990, \, e_1 =e_2 =e_3 \approx -1.076, \, e_0 =0, \\
p_1 &= p_2 =0, \, p_3 = p_4 \approx -1.310, \, r_{12} \approx .6215,\, r_{21} \approx .5004, \, r_{22} \approx -.02231, \, r_{32} \approx -.1930,\\
\xi^1 &\approx -.1504,\,\xi^2 \approx 2.682, \, \xi^3 + \xi^4 \approx -2.573, \, u^1 =u^2 = u^3 \approx 1.336,\\
e^D &\approx .3481, \, \mathcal{Z}^1 \approx .1845, \, \mathcal{Z}^2 \approx 2.333, \, \mathcal{Z}^3 = \mathcal{Z}^4 \approx .3810, \, v^1 \approx 2.202, \, v^2 \approx 18.73, \, v^3 \approx 4.023.
\end{align*}
Since we have $\e \approx 0$ this corresponds to a de Sitter extremum. Calculating the $\eta$ parameter for this solution we find $\eta \approx -3.7$. So this solution is not a minimum but rather a saddle point. From the mass matrix for the moduli one sees that there is exactly one tachyonic direction that is a mixture of several moduli including the axions. We have looked at several extremal points for this model but always found at least one tachyonic direction with $\eta \lesssim -2.4$.
We did not pursue this model further since it is not compatible with the base-fiber construction mentioned above. Splitting the compact space into a base and a fiber always results in an $r$ matrix that has one row equal to zero. Due to the factorization in the K\"ahler sector we therefore find $\e \geq \frac95$ for all models that can be obtained from the base-fiber construction.
In a related work \cite{Caviezel:2008tf} the authors searched for slow-roll inflation and de Sitter vacua in coset spaces \cite{Tomasiello:2007eq}, \cite{Koerber:2008rx}, \cite{Caviezel:2008ik}. They found that for an orientifold of SU(2)$\times$SU(2) one can obtain de Sitter extrema with one tachyonic direction. This orientifold of SU(2)$\times$SU(2) can be thought of as a $\Z_2\times\Z_2$ quotient of $T^6$ with metric fluxes as was discussed in \cite{Aldazabal:2007sn}. This means that at least a subset of the compact spaces exists although it is not possible to obtain them from the base-fiber construction. The authors of \cite{Caviezel:2008tf} also checked whether the no-go theorems related to the $\eta$ parameter \cite{Covi:2008ea}, \cite{Covi:2008cn} can be applied to their SU(2)$\times$SU(2) orientifold but found that this is not the case.
It would be interesting to study this model further to verify whether all solutions to the Bianchi identities that give small $\e$ have a corresponding compact space and whether it is possible to find de Sitter vacua that have no tachyonic directions.

\subsection{Non-standard orientifold of $\Z_2\times\Z_2$}\label{nonstandardZ2}

For the non-standard orientifold projection we can explicitly construct solutions to the Bianchi identities that lead to vanishing $\e$. The two interesting cases have the 2 dimensional submanifolds spanned by the 3 and 5 or 4 and 6 directions as base and the other four transverse directions as fiber.
The first case leads to $p_1, p_4, r_{11}, r_{14}, r_{23}, \hat{r}^2_1$ fluxes with all other NSNS fluxes equal to zero. The non-zero fluxes have to satisfy the Bianchi identity  $2 r_{11} r_{23}-r_{14} \hat{r}^2_1=0$.
The second case with the 4 and 6 direction as base allows for $p_1, p_4, r_{11}, r_{14}, r_{22}, \hat{r}^3_1$ fluxes with all other NSNS fluxes equal to zero. The non-zero fluxes have to satisfy the Bianchi identity  $2 r_{11} r_{22}-r_{14} \hat{r}^3_1=0$. This case is related to the first one by symmetry so we will only focus on the first case with the 3 and 5 direction as base.
In the first case we can solve the remaining Bianchi identity $2 r_{11} r_{23}=r_{14} \hat{r}^2_1$ by setting one of the metric fluxes appearing on either side to zero. If we have $r_{23} =\hat{r}^2_1=0$ or $r_{11} = r_{14} =0$ our no-go theorems based on the factorization in the K\"ahler sector give $\e \geq 2$ and $\e \geq \frac95$, respectively. The other two possibilities $r_{11} = \hat{r}^2 =0$ and $r_{23} = r_{14} =0$ give numerically $\e \approx \frac43$ and $\e \approx .2$, so that we focus on solutions that have $2 r_{11} r_{23}=r_{14} \hat{r}^2_1 \neq 0$. For $2 r_{11} r_{23}=r_{14} \hat{r}^2_1 < 0$ we find numerically that $\e \geq .2$ where the lower bound is attained in the limit where $r_{23} = r_{14} \approx 0$.
So the only solution to the Bianchi identities that leads to vanishing $\e$ is $2 r_{11} r_{23}=r_{14} \hat{r}^2_1 > 0$. The quantization condition in this case forces $2 r_{11} r_{23}=r_{14} \hat{r}^2_1 = n^2 \pi^2, \, n \in \mathbb{Z}$. There are two different solutions. For $n=2k$ even we find
\begin{align}
r_{14} = \frac{n_1}{n_2} r_{11}, \, \hat{r}^2_1 =\frac{4 k^2 \pi^2 n_2}{n_1 r_{11}},\, r_{23} = \frac{2 k^2 \pi^2}{r_{11}}, \, p_1 = n_2 \left( 12 \sqrt{2} +\frac{p_4}{n_1} \right), \quad k,n_1, n_2 \in \mathbb{Z}.
\end{align}
For $n= 2k+1$ odd we find
\begin{align}
r_{14} &= \pm r_{11}, \, \hat{r}^2_1 =\pm \frac{(2 k+1)^2 \pi^2}{r_{11}},\, r_{23} = \frac{(2 k+1)^2 \pi^2}{2 r_{11}}, \\
p_1 &= 6 \left( \sqrt{2} n_1 + n_2 r_{11} \right), p_4 = \mp 6\left(\sqrt{2} n_1 - n_2 r_{11} \right), \quad k,n_1, n_2 \in \mathbb{Z}.
\end{align}
Note that not all metric fluxes are quantized. $r_{11}$ can take arbitrary values. Both solutions respect the symmetry arising from shifting the $B$ field and $H$ flux. Under a shift of $u^1 \rightarrow u^1 + a^1$ we have $p_1 \rightarrow r_{11} a^1$ and $p_4 \rightarrow r_{14} a^1$ so that we can set one of $p_1$ and $p_4$ equal to zero. We will set $p_4=0$ and minimize $\e$ numerically for integers $k,n_1, n_2 \in \mathbb{Z}$. One particular solution for $n=2k$ even with $\e \approx 10^{-19}$ is\footnote{The following values are rounded to six digits and give $\e \approx 10^{-4}$. Note also, that similar to the previous case, the $C_3$ axions $\xi^K$ appear in the potential only through the linear combinations $p_K \xi^K$ and $r_{aK} \xi^K$. Since for this model $r$ has two rows we can stabilize only three linear combinations of them. In particular for $p_4=0$ we can stabilize $p_1 \xi^1$, $r_{11} \xi^1 +r_{14} \xi^4$ and $r_{23} \xi^3$. This means that $\xi^2$ remains unstabilized.}
\begin{align*}
\widetilde{m} &\approx -3.74854, \, m_1 \approx -32.5482, \, m_2 \approx -22.5086, \\
e_1 &\approx 2.76717,\, e_2 \approx -2.92192, \, e_0 \approx -.251057, \\
k &= 1,\, n_1= -3, \, n_2 = -1, \, r_{11} \approx -1.62809,\\
\xi^1 &\approx -6.39013,\, \xi^2 \approx \text{unstabilized}, \, \xi^3 \approx -1.66584, \, \xi^4 \approx -15.7204,\\
u^1 &\approx -2.49321,\, u^2 \approx 3.16322, \, v^1 \approx 3.32339, \, v^2 \approx 11.6507,\\
e^D &\approx .0745145, \, \mathcal{Z}^1 \approx .413947, \, \mathcal{Z}^2 \approx 38.0222, \, \mathcal{Z}^3 \approx .360619,\, \mathcal{Z}^4 \approx 3.65332.
\end{align*}
This particular solution has one tachyonic direction and $\eta \approx -2.5$. The tachyon is a mixture of several moduli including the axions.
We have scanned over ranges where the flux quanta $n_1,n_2,k$ are of order 1 for both $n$ even and odd and found dozens of solutions. All of these solutions had at least one tachyonic direction that generically is a mixture of all moduli. We generically found that $\eta \lesssim -2.4$ and solutions close to that bound have only one single tachyonic directions. The no-go theorems of \cite{Covi:2008ea}, \cite{Covi:2008cn} cannot be applied to this particular model since we have D-terms. It would be very interesting to understand this tachyonic direction that appears in both of the models in this section better. We, of course, cannot rule out that there are solutions corresponding to metastable de Sitter vacua since we only did a numerical study but due to the large number of solutions that all have this tachyonic direction with roughly the same value for $\eta$ that furthermore is independent of fluxes, we suspect that this model has no metastable de Sitter vacua.
We have examined the vicinity of our extrema in which $\e$ is still small to see whether this enables us to find small $|\eta|$ to satisfy the conditions for slow-roll inflation, but we found that $\eta$ changes at most by a factor of two in this region. We have also minimized $\epsilon$ with constraints ensuring small $|\eta|$, but have not been able to find small $\epsilon$ in this case. We take this as a strong indication that these models are incompatible with slow-roll inflation. However, we do not have an analytic proof of this, and it would be very interesting to investigate this further. We will leave this for future research.

\section{Conclusions}
We have explored the possibility of slow-roll inflation and de Sitter vacua in type IIA compactifications that include standard NSNS 3-form fluxes, RR fluxes, D6-branes and O6-planes as well as metric fluxes. We have derived a set of no-go theorems based on the dependence of the potential on the dilaton, volume, K\"{a}hler and complex structure moduli, extending previous work by HKTT \cite{Hertzberg:2007wc}. Theorems of this kind are valuable because they specify the minimal set of ingredients required to have slow-roll inflation or de Sitter vacua in this type of compactifications, or put differently they rule out entire regions in the vast landscape of solutions of string theory. To demonstrate their usefulness, we applied these no-go theorems to toroidal orientifolds with abelian orbifold groups generated by rotations and reflections, that, in the absence of fluxes and after orientifolding, preserve $\mathcal{N}=1$ supersymmetry. As we showed, the application of the no-go theorems is straightforward in some cases while in others T-dualities and field redefinitions play a crucial role. We find that under the assumptions made in deriving the no-go theorems, the slow-roll parameter $\epsilon$ is bounded from below in all models of this class except the two $\mathbb{Z}_2\times\mathbb{Z}_2$ cases. In those cases, we have succeeded in finding regions of parameter space where the slow-roll parameter $\epsilon$ is very small numerically, but unfortunately $\eta$ turns out to be such that inflation is much too short, making these compactifications uninteresting from a cosmological perspective.

While it would be more satisfying and insightful to have no-go theorems for $\epsilon$ and $\eta$ simultaneously, the ones obtained in this paper are exclusively for $\epsilon$. Our exploration of the range of $\eta$ has always been numerical. We either computed $\eta$ where $\epsilon$ had already been found to be small or have failed to find a small value for $\epsilon$ when we restricted the minimization procedure to keeping $\eta$ small. Thus, although we are confident of our results we lack the insight as to why the necessary conditions for small $\eta$ are not compatible with those for small  $\epsilon$.

There are several effects we have not considered. We have ignored twisted sector modes and blow-up modes. We have also ignored more general brane configurations such as backreacting D6-branes that do not wrap rigid cycles and are far from their static configuration, coisotropic branes, or NSNS sources. Even though this is by no means guaranteed, all of these ingredients might render the no-go theorems invalid and may be interesting to investigate further. We leave this for future work.

There is an orthogonal line of research pursued in~\cite{Steinhardt:2008nk} that comes to similar conclusions. While we have not had the chance to do so, we think it would be interesting to understand if there is a relation between the two.

\section*{Acknowledgements}

We would like to thank Davide Cassani, Thomas Grimm, Luca Martucci, Erik Plauschinn, Eva Silverstein, Paul Steinhardt, Dimitrios Tsimpis, Alexander Westphal, and in particular Matthias Ihl for discussion during the early stages of this project, as well as Claudio Caviezel,  Paul Koerber, Simon K{\"o}rs, Dieter L{\"u}st and Marco Zagermann for useful discussions and correspondence.
S. Paban and D. Robbins would like to thank the Aspen Center for Physics for hospitality while part of this work was done, and R. Flauger would like to thank the Princeton Center for Theoretical Sciences for hospitality during the final stages of this work.  T. Wrase is supported by the German Research Foundation (DFG) within the Emmy-Noether-Program (Grant number ZA 279/1-2).
The work of R. Flauger, S. Paban and D. Robbins has been partially supported by the National Science Foundation under Grant No. PHY-0455649.
\appendix

\section{Conventions}
\label{sec:Conventions}
Our conventions largely follow \cite{Grimm:2004ua}.

Consider type IIA string theory on a Calabi-Yau three-fold $X$, equipped with a $\Z_2$ orientifold action which includes an anti-holomorphic involution $\s$.  The cohomology of $X$ then splits into even and odd parts, depending upon the behavior of each class under $\s$.  We will take the following basis of representative forms:
\begin{itemize}
\item The zero-form 1,
\item a set of odd two-forms $\om_a$, $a=1,\ldots,h^{1,1}_-$,
\item a set of even two-forms $\m_\al$, $\al=1,\ldots,h^{1,1}_+$,
\item a set of even four-forms $\widetilde\om^a$, $a=1,\ldots,h^{1,1}_-$,
\item a set of odd four-forms $\widetilde\m^\al$, $\al=1,\ldots,h^{1,1}_+$,
\item a six form $\varphi$, odd under $\s$,
\item a set of even three-forms $a_K$, $K=1,\ldots,h^{2,1}+1$,
\item and a set of odd three-forms $b^K$, $K=1,\ldots,h^{2,1}+1$.
\end{itemize}
Additionally, it turns out that we can always choose the $a_K$ and $b^K$ to form a symplectic basis such that the only non-vanishing intersections are
\be
\int_X a_K\w b^J=\d_K^J.
\ee

Similarly, we can take the even-degree forms to obey
\be \label{eq:tripleintersect}
\int_X\varphi = 1,\qquad\int_X\om_a\w\om_b\w\om_c=\k_{abc},\qquad\int_X\om_a\w\m_\al\w\m_\beta=\widehat{\k}_{a\,\al\beta},\non
\ee
\be
\int_X\om_a\w\widetilde{\om}^b=\d_a^b,\qquad\int_X\m_\al\w\widetilde{\m}^\beta=\d_\al^\beta.
\ee

Now let us describe the four-dimensional fields of this class of compactifications, restricting ourselves, for simplicity, to the bosonic sector.  First we have the K\"ahler moduli, parametrized by complex scalar fields $t^a=u^a+iv^a$ coming from the expansion
\be
B+iJ=J_c=t^a\om_a,
\ee
where the complexified K\"ahler form $J_c$ must be odd under $\s$.  Note that the K\"ahler form $J=v^a\om_a$ determines the compactification volume (in string frame) via
\be
\mathcal{V}_6=\frac{1}{3!}\int_XJ\w J\w J=\frac{1}{6}\k_{abc}v^av^bv^c.
\ee

To describe the complex moduli, let us write the holomorphic three-form as
\be
\Om=\mathcal{Z}^Ka_K-\mathcal{F}_Kb^K.
\ee
We will use conventions in which
\be
i\int_X\Om\w\overline{\Om}=1,\qquad\s^\ast\Om=\overline{\Om},
\ee
so that the $\mathcal{Z}^K$ are real functions of the complex moduli and $\mathcal{F}_K$ are pure imaginary, and together they satisfy the constraint $\mathcal{Z}^K\mathcal{F}_K=-i/2$.  We can now define a complexified version~\cite{Grimm:2004ua}
\be
\Om_c=C_3+2ie^{-D}\Re\Om=\lp\xi^K+2ie^{-D}\mathcal{Z}^K\rp a_K,
\ee
where $e^{-D}=\mathcal{V}_6^{1/2}e^{-\phi}$ contains the dilaton and we expand the periods of $C_3$ (which must be even under $\s$ in order to survive the orientifold projection) as $C_3=\xi^Ka_K$.  Note that we abuse notation somewhat here as we ignore other pieces which contribute to the ten-dimensional RR three-form potential $C_3$, namely pieces that give rise to four-dimensional vectors and (local) pieces that give the four-form RR  flux, both of which will be discussed below.  The complex moduli $N^K=\hlf\xi^K+ie^{-D}\mathcal{Z}^K$ are then simply given by the expansion
\be
\Om_c=2N^Ka_K,
\ee
and include the complex structure moduli of the metric, the dilaton, and the RR three-form periods.

There are also $h^{1,1}_+$ four-dimensional vectors from the decomposition of the RR three-form potential, which includes a contribution
\be
C_3=A^\al\w\m_\al.
\ee

We can now consider turning on fluxes.  In the RR sector, this leads us to include
\be
F_0=\widetilde{m},\qquad F_2=m^a\om_a,\qquad F_4=e_a\widetilde{\om}^a,\qquad F_6=\widetilde{e}\varphi.
\ee
From the NSNS sector, we can include the usual $H$-flux,
\be
H=p_Kb^K,
\ee
but we can also consider generalized metric fluxes.  For more details, please refer to sections \ref{subsec:generalized} and \ref{subsec:fluxes}, but for completeness we list the definitions of our parameters $r_{aK}$ and $\hat{r}_\al^K$,
\be
d\om_a=-r_{aK}b^K,\qquad d\mu_\al=-\hat{r}_\al^Ka_K.
\ee


\begin{thebibliography}{10}

\bibitem{Baumann:2008aq}
D.~Baumann {\em et al.}, ``{CMBPol Mission Concept Study: Probing Inflation
  with CMB Polarization},''
\href{http://arXiv.org/abs/0811.3919}{{\tt 0811.3919}}.

\bibitem{McAllister:2007bg}
L.~McAllister and E.~Silverstein, ``{String Cosmology: A Review},'' {\em Gen.
  Rel. Grav.} {\bf 40} (2008) 565--605,
\href{http://arXiv.org/abs/0710.2951}{{\tt 0710.2951}}.

\bibitem{Kachru:2003aw}
S.~Kachru, R.~Kallosh, A.~Linde, and S.~P. Trivedi, ``{De Sitter vacua in
  string theory},'' {\em Phys. Rev.} {\bf D68} (2003) 046005,
\href{http://arXiv.org/abs/hep-th/0301240}{{\tt hep-th/0301240}}.

\bibitem{Quevedo:2002xw}
F.~Quevedo, ``{Lectures on string/brane cosmology},'' {\em Class. Quant. Grav.}
  {\bf 19} (2002) 5721--5779,
\href{http://arXiv.org/abs/hep-th/0210292}{{\tt hep-th/0210292}}.

\bibitem{Linde:2005dd}
A.~Linde, ``{Inflation and string cosmology},'' {\em ECONF} {\bf C040802}
  (2004) L024,
\href{http://arXiv.org/abs/hep-th/0503195}{{\tt hep-th/0503195}}.

\bibitem{HenryTye:2006uv}
S.~H. Henry~Tye, ``{Brane inflation: String theory viewed from the cosmos},''
  {\em Lect. Notes Phys.} {\bf 737} (2008) 949--974,
\href{http://arXiv.org/abs/hep-th/0610221}{{\tt hep-th/0610221}}.

\bibitem{Cline:2006hu}
J.~M. Cline, ``{String cosmology},''
\href{http://arXiv.org/abs/hep-th/0612129}{{\tt hep-th/0612129}}.

\bibitem{Burgess:2007pz}
C.~P. Burgess, ``{Lectures on Cosmic Inflation and its Potential Stringy
  Realizations},'' {\em PoS} {\bf P2GC} (2006) 008,
\href{http://arXiv.org/abs/0708.2865}{{\tt 0708.2865}}.

\bibitem{Kallosh:2007ig}
R.~Kallosh, ``{On Inflation in String Theory},'' {\em Lect. Notes Phys.} {\bf
  738} (2008) 119--156,
\href{http://arXiv.org/abs/hep-th/0702059}{{\tt hep-th/0702059}}.

\bibitem{DeWolfe:2005uu}
O.~DeWolfe, A.~Giryavets, S.~Kachru, and W.~Taylor, ``{Type IIA moduli
  stabilization},'' {\em JHEP} {\bf 07} (2005) 066,
\href{http://arXiv.org/abs/hep-th/0505160}{{\tt hep-th/0505160}}.

\bibitem{Banks:2006hg}
T.~Banks and K.~van~den Broek, ``{Massive IIA flux compactifications and
  U-dualities},'' {\em JHEP} {\bf 03} (2007) 068,
\href{http://arXiv.org/abs/hep-th/0611185}{{\tt hep-th/0611185}}.

\bibitem{Hertzberg:2007ke}
M.~P. Hertzberg, M.~Tegmark, S.~Kachru, J.~Shelton, and O.~\"Ozcan, ``{Searching
  for Inflation in Simple String Theory Models: An Astrophysical
  Perspective},'' {\em Phys. Rev.} {\bf D76} (2007) 103521,
\href{http://arXiv.org/abs/0709.0002}{{\tt 0709.0002}}.

\bibitem{Hertzberg:2007wc}
M.~P. Hertzberg, S.~Kachru, W.~Taylor, and M.~Tegmark, ``{Inflationary
  Constraints on Type IIA String Theory},'' {\em JHEP} {\bf 12} (2007) 095,
\href{http://arXiv.org/abs/0711.2512}{{\tt 0711.2512}}.

\bibitem{Villadoro:2005cu}
G.~Villadoro and F.~Zwirner, ``{N = 1 effective potential from dual type-IIA
  D6/O6 orientifolds with general fluxes},'' {\em JHEP} {\bf 06} (2005) 047,
\href{http://arXiv.org/abs/hep-th/0503169}{{\tt hep-th/0503169}}.

\bibitem{Ihl:2006pp}
M.~Ihl and T.~Wrase, ``{Towards a realistic type IIA T**6/Z(4) orientifold
  model with background fluxes. I: Moduli stabilization},'' {\em JHEP} {\bf 07}
  (2006) 027,
\href{http://arXiv.org/abs/hep-th/0604087}{{\tt hep-th/0604087}}.

\bibitem{Silverstein:2007ac}
E.~Silverstein, ``{Simple de Sitter Solutions},'' {\em Phys. Rev.} {\bf D77}
  (2008) 106006,
\href{http://arXiv.org/abs/0712.1196}{{\tt 0712.1196}}.

\bibitem{Silverstein:2008sg}
E.~Silverstein and A.~Westphal, ``{Monodromy in the CMB: Gravity Waves and
  String Inflation},'' {\em Phys. Rev.} {\bf D78} (2008) 106003,
\href{http://arXiv.org/abs/0803.3085}{{\tt 0803.3085}}.

\bibitem{Haque:2008jz}
S.~S. Haque, G.~Shiu, B.~Underwood, and T.~Van~Riet, ``{Minimal simple de
  Sitter solutions},''
\href{http://arXiv.org/abs/0810.5328}{{\tt 0810.5328}}.

\bibitem{Palti:2008mg}
  E.~Palti, G.~Tasinato and J.~Ward,
  ``WEAKLY-coupled IIA Flux Compactifications,''
  JHEP {\bf 0806}, 084 (2008)
  [arXiv:0804.1248 [hep-th]].

\bibitem{Caviezel:2008tf}
C.~Caviezel, P.~Koerber, S.~K\"ors, D.~L\"ust, T.~Wrase and M.~Zagermann, ``{On the Cosmology of Type IIA Compactifications on
  SU(3)- structure Manifolds},''
\href{http://arXiv.org/abs/0812.3551}{{\tt 0812.3551}}.

\bibitem{Scherk:1979zr}
J.~Scherk and J.~H. Schwarz, ``How to get masses from extra dimensions,'' {\em
  Nucl. Phys.} {\bf B153} (1979)
61--88.

\bibitem{Kaloper:1998kr}
N.~Kaloper, R.~R. Khuri, and R.~C. Myers, ``On generalized axion reductions,''
  {\em Phys. Lett.} {\bf B428} (1998) 297--302,
\href{http://arXiv.org/abs/hep-th/9803066}{{\tt hep-th/9803066}}.

\bibitem{Kaloper:1999yr}
N.~Kaloper and R.~C. Myers, ``The {O}(dd) story of massive supergravity,'' {\em
  JHEP} {\bf 05} (1999) 010,
\href{http://arXiv.org/abs/hep-th/9901045}{{\tt hep-th/9901045}}.

\bibitem{Ihl:2007ah}
M.~Ihl, D.~Robbins, and T.~Wrase, ``{Toroidal Orientifolds in IIA with General
  NS-NS Fluxes},'' {\em JHEP} {\bf 08} (2007) 043,
\href{http://arXiv.org/abs/0705.3410}{{\tt 0705.3410}}.

\bibitem{Robbins:2007yv}
D.~Robbins and T.~Wrase, ``{D-Terms from Generalized NS-NS Fluxes in Type
  II},'' {\em JHEP} {\bf 12} (2007) 058,
\href{http://arXiv.org/abs/0709.2186}{{\tt 0709.2186}}.

\bibitem{Hellerman:2002ax}
S.~Hellerman, J.~McGreevy, and B.~Williams, ``{Geometric Constructions of
  Nongeometric String Theories},'' {\em JHEP} {\bf 01} (2004) 024,
\href{http://arXiv.org/abs/hep-th/0208174}{{\tt hep-th/0208174}}.

\bibitem{Dabholkar:2002sy}
A.~Dabholkar and C.~Hull, ``{Duality twists, orbifolds, and fluxes},'' {\em
  JHEP} {\bf 09} (2003) 054,
\href{http://arXiv.org/abs/hep-th/0210209}{{\tt hep-th/0210209}}.

\bibitem{Shelton:2005cf}
J.~Shelton, W.~Taylor, and B.~Wecht, ``{Nongeometric Flux Compactifications},''
  {\em JHEP} {\bf 10} (2005) 085,
\href{http://arXiv.org/abs/hep-th/0508133}{{\tt hep-th/0508133}}.

\bibitem{Shelton:2006fd}
J.~Shelton, W.~Taylor, and B.~Wecht, ``{Generalized flux vacua},'' {\em JHEP}
  {\bf 02} (2007) 095,
\href{http://arXiv.org/abs/hep-th/0607015}{{\tt hep-th/0607015}}.

\bibitem{Aldazabal:2007sn}
G.~Aldazabal and A.~Font, ``{A second look at N=1 supersymmetric AdS4 vacua of
  type IIA supergravity},'' {\em JHEP} {\bf 02} (2008) 086,
\href{http://arXiv.org/abs/0712.1021}{{\tt 0712.1021}}.

\bibitem{Wecht:2007wu}
B.~Wecht, ``{Lectures on Nongeometric Flux Compactifications},'' {\em Class.
  Quant. Grav.} {\bf 24} (2007) S773--S794,
\href{http://arXiv.org/abs/0708.3984}{{\tt 0708.3984}}.

\bibitem{Grimm:2004ua}
T.~W. Grimm and J.~Louis, ``{The effective action of type IIA Calabi-Yau
  orientifolds},'' {\em Nucl. Phys.} {\bf B718} (2005) 153--202,
\href{http://arXiv.org/abs/hep-th/0412277}{{\tt hep-th/0412277}}.

\bibitem{Grana:2006kf}
M.~Gra\~{n}a, R.~Minasian, M.~Petrini, and A.~Tomasiello, ``{A scan for new N=1
  vacua on twisted tori},'' {\em JHEP} {\bf 05} (2007) 031,
\href{http://arXiv.org/abs/hep-th/0609124}{{\tt hep-th/0609124}}.

\bibitem{Erler:1992ki}
J.~Erler and A.~Klemm, ``{Comment on the generation number in orbifold
  compactifications},'' {\em Commun. Math. Phys.} {\bf 153} (1993) 579--604,
\href{http://arXiv.org/abs/hep-th/9207111}{{\tt hep-th/9207111}}.

\bibitem{Reffert:2006du}
S.~Reffert, ``{Toroidal orbifolds: Resolutions, orientifolds and applications
  in string phenomenology},''
\href{http://arXiv.org/abs/hep-th/0609040}{{\tt hep-th/0609040}}.

\bibitem{Reffert:2007im}
  S.~Reffert,
  ``The Geometer's Toolkit to String Compactifications,''
  arXiv:0706.1310 [hep-th].

\bibitem{Villadoro:2007tb}
G.~Villadoro and F.~Zwirner, ``{On general flux backgrounds with localized
  sources},'' {\em JHEP} {\bf 11} (2007) 082,
\href{http://arXiv.org/abs/0710.2551}{{\tt 0710.2551}}.

\bibitem{Tomasiello:2007eq}
A.~Tomasiello, ``{New string vacua from twistor spaces},'' {\em Phys. Rev.}
  {\bf D78} (2008) 046007,
\href{http://arXiv.org/abs/0712.1396}{{\tt 0712.1396}}.

\bibitem{Koerber:2008rx}
P.~Koerber, D.~L\"ust, and D.~Tsimpis, ``{Type IIA AdS4 compactifications on
  cosets, interpolations and domain walls},'' {\em JHEP} {\bf 07} (2008) 017,
\href{http://arXiv.org/abs/0804.0614}{{\tt 0804.0614}}.

\bibitem{Caviezel:2008ik}
C.~Caviezel, P.~Koerber, S.~K\"ors, D.~L\"ust, D.~Tsimpis and M.~Zagermann, ``{The effective theory of type IIA AdS4
  compactifications on nilmanifolds and cosets},''
\href{http://arXiv.org/abs/0806.3458}{{\tt 0806.3458}}.

\bibitem{Bergman:2007qq}
A.~Bergman and D.~Robbins, ``{Ramond-Ramond Fields, Cohomology and
  Non-Geometric Fluxes},''
\href{http://arXiv.org/abs/0710.5158}{{\tt 0710.5158}}.

\bibitem{Covi:2008ea}
L.~Covi, M.~Gomez-Reino, C.~Gross, J.~Louis, G.~A.~Palma and C.~A.~Scrucca, ``{de Sitter vacua in no-scale supergravities and
  Calabi-Yau string models},'' {\em JHEP} {\bf 06} (2008) 057,
\href{http://arXiv.org/abs/0804.1073}{{\tt 0804.1073}}.

\bibitem{Covi:2008cn}
L.~Covi, M.~Gomez-Reino, C.~Gross, J.~Louis, G.~A.~Palma and C.~A.~Scrucca, ``{Constraints on modular inflation in supergravity and
  string theory},'' {\em JHEP} {\bf 08} (2008) 055,
\href{http://arXiv.org/abs/0805.3290}{{\tt 0805.3290}}.

\bibitem{Steinhardt:2008nk}
P.~J. Steinhardt and D.~Wesley, ``{Dark Energy, Inflation and Extra
  Dimensions},''
\href{http://arXiv.org/abs/0811.1614}{{\tt 0811.1614}}.

\end{thebibliography}
\providecommand{\href}[2]{#2}\begingroup\raggedright\endgroup

\end{document}